\documentclass[11pt,a4paper]{article}
\pdfoutput=1
\usepackage{afterpage}
\usepackage{varwidth}
\usepackage{tikz}
\usetikzlibrary{shapes}
\usepackage{relsize}
\usepackage{subfigure}
\makeatletter
\newcommand\mathcircled[1]{%
  \mathpalette\@mathcircled{#1}%
}
\newcommand\@mathcircled[2]{%
  \tikz[baseline=(math.base)] \node[draw,ellipse,inner sep=1pt] (math) {$\m@th#1#2$};%
}
\makeatother
\usepackage{longtable}
\usepackage[new]{old-arrows}
\usepackage{amsmath}
\usepackage{lscape}
\usepackage{caption}
\usepackage{jheppub}
\usepackage[utf8]{inputenc}
\usepackage{soul,xcolor}
\usepackage{cancel}
\usepackage{makecell}
\usepackage{diagbox}
\usepackage{makecell}
\usepackage[normalem]{ulem}
\newsavebox{\foobox}

\usepackage{multirow}
\usepackage{dcolumn}
\newcommand\Tstrut{\rule{0pt}{2.6ex}}         
   
\usepackage{ulem}
\usepackage{graphicx}
\usepackage{comment}
\usepackage{bm,array}
\usepackage{graphics}
\usepackage{epsf}
\usepackage{color}
\usepackage{amsmath}
\usepackage{amssymb}
\usepackage{latexsym}
\usepackage{slashed}
\usepackage{float}
\usepackage{cases}
\usepackage{multirow}
\usepackage{framed}
\def\be{\begin{equation}}
\def\ee{\end{equation}}
\def\ba{\begin{array}}
\def\ea{\end{array}}
\usepackage{amsmath}
\usepackage{amssymb}
\usepackage[utf8]{inputenc}

\def\alambda{A_\lambda}
\def\akappa{A_\kappa}
\def\mueff{\mu_\mathrm{eff}}
\def\beff{B_{\mathrm{eff}}}
\def\tanb{\tan\beta}

\def\wpm{W^\pm}

\def\bbar{\bar{b}}
\def\tbar{\bar{t}}
\def\ttbar{t\bar{t}}
\def\bbbar{b\bar{b}}

\def\sQ3{\widetilde{Q}_3}
\def\sU3{\widetilde{U}_3}
\def\sD3{\widetilde{D}_3}
\def\stleft{\tilde{t}_L}
\def\stright{\tilde{t}_R}
\def\sbleft{\tilde{b}_L}
\def\sbright{\tilde{b}_R}
\def\stone{\tilde{t}_1}

\def\stonetwo{\tilde{t}_{1,2}}

\def\hsm{h_{\rm SM}}
\def\hpm{H^\pm}
\def\hs{h_{_S}} 
\def\as{a_{_S}}
\def\bino{\widetilde{B}}
\def\wino{\widetilde{W}}

\def\higgsinod{\widetilde{H}^0_d}
\def\higgsinou{\widetilde{H}^0_u}
\def\singlino{\widetilde{S}}
\def\ntrli{\chi_{_i}^0}
\def\ntrlj{\chi_{_j}^0}
\def\ntrlk{\chi_{_k}^0}
\def\ntrlone{\chi_{_1}^0}
\def\ntrltwo{\chi_{_2}^0}
\def\ntrlthree{\chi_{_3}^0}
\def\ntrlfour{\chi_{_4}^0}
\def\ntrlfive{\chi_{_5}^0}
\def\ntrlonetwo{\chi_{_{1,2}}^0}

\def\ntrlthreefour{\chi_{_{3,4}}^0}

\def\charjmp{\chi_j^\mp}
\def\charonep{\chi_{_1}^+}

\def\charonepm{\chi_{_1}^\pm}

\def\chartwopm{\chi_{_2}^\pm}

\def\mone{M_1}
\def\mtwo{M_2}

\def\mhs{m_{_{h_S}}}
\def\mhssq{m^2_{_{h_S}}}
\def\mas{m_{a_{_S}}}
\def\massq{m^2_{a_{_S}}}

\def\mstone{m_{\tilde{t}_1}}

\def\thetastop{\theta_{\tilde{t}}}
\def\costhetastop{\cos\theta_{\tilde{t}}}
\def\sinthetastop{\sin\theta_{\tilde{t}}}
\def\msinglino{m_{_{\widetilde{S}}}}
\def\msinglinosq{m^2_{_{\widetilde{S}}}}
\def\mntrli{m_{{_{\chi}}_i^0}}

\def\mntrlone{m_{{_{\chi}}_{_1}^0}}
\def\mntrltwo{m_{{_{\chi}}_{_2}^0}}

\def\mntrlfive{m_{{_{\chi}}_{_5}^0}}

\def\mntrlthreefour{m_{{_{\chi}}_{_{3,4}}^0}}
\def\mntrlonetwo{m_{{_{\chi}}_{_{1,2}}^0}}
\def\mcharone{m_{{_{\chi}}_{_1}^\pm}}
\def\mchartwo{m_{{_{\chi}}_{_2}^\pm}}
\def\mhsm{m_{h_{\mathrm{SM}}}}

\newcommand{\vu}{v_u}
\newcommand{\vd}{v_d}
\newcommand{\vs}{v_s}
\def\vev{{\it vev}}
\def\vevs{{\it vevs}}
\def\vu{v_u}
\def\vd{v_d}

\def\vs{v_{\!_S}}

\def\pt{p_{_T}}
\def\ptell{\pt^{(\ell)}}

\def\ptjet{\pt^{(j)}}
\def\etaell{\eta_{_{\ell}}}
\def\deltar{\Delta R}

\def \MET{E{\!\!\!/}_T}
\def\beq{\begin{equation}}
\def\eeq{\end{equation}}
\def\beqa{\begin{eqnarray}}
\def\eeqa{\end{eqnarray}}

\allowdisplaybreaks
\newcommand{\fbinv}{\text{fb}$^{-1}$}

\def\etmiss{\slashed{E}_T}

\def\nmssmtools{{\tt NMSSMTools}}

\def\micromegas{{\tt micrOMEGAs}}
\def\higgsbounds{{\tt HiggsBounds}}
\def\higgssignals{{\tt HiggsSignals}}

\def\pythia8{{\tt PYTHIA8}}

\def \MET{E{\!\!\!/}_T}
\def\br {\begin{eqnarray}}
\def\er {\end{eqnarray}}
\def\z3nmssm{$Z_3$-NMSSM}
\newcommand{\cred}[1]{{\color{red}#1}}
\newcommand{\cblue}[1]{{\color{blue}#1}}

\newcommand{\cmagenta}[1]{{\color{magenta}#1}}

\newcommand{\bea}{\begin{eqnarray}}
\newcommand{\eea}{\end{eqnarray}}

\title{Hunting ewinos and a light scalar of $Z_3$-NMSSM with a bino-like dark matter in top squark decays at the LHC}
\author[a]{AseshKrishna Datta,}
\author[b]{Monoranjan Guchait,}
\author[b]{Arnab Roy}
\author[a,c]{and Subhojit Roy}
\affiliation[a]{Harish-Chandra Research Institute, A CI of Homi Bhabha National
Institute, Chhatnag Road, Jhunsi, Prayagraj (Allahabad) 211019, India}
\affiliation[b]{Department of High Energy Physics, Tata Institute of Fundamental Research, 
Homi Bhabha Road, Mumbai-400005, India}
\affiliation[c]{Regional Centre for Accelerator-based Particle Physics, Harish-Chandra 
Research Institute, \\Chhatnag Road, Jhunsi, Prayagraj (Allahabad) 211019, India}
\emailAdd{asesh@hri.res.in, guchait@tifr.res.in, arnab.roy@tifr.res.in, 
subhojitroy@hri.res.in}
\preprint{HRI-RECAPP-2022-014}
\abstract{
We explore a possible signal of the presence of relatively light electroweakinos (ewinos) and a singlet-like scalar ($\as$) of the 
$Z_3$-symmetric Next-to-Minimal Supersymmetric Standard Model (\z3nmssm) in
the cascade decays of not so heavy top squarks ($\mstone~\lesssim~1.5$~TeV) that may be produced in pairs at the Large Hadron Collider LHC.
We work in a scenario where the lightest (next-to-lightest) SUSY particle is 
bino (singlino)-like with its mass below 100 GeV and is mildly tempered with higgsino and singlino admixtures. The singlet-like scalar provides an annihilation funnel for the bino-like states such that the latter could act as a viable dark matter candidate, unlike what is now highly constrained in the MSSM.
We consider a pair of immediately heavier neutralinos and the lighter chargino which all are higgsino-like with masses in the range $\sim 0.5$--$1$~TeV and are still compatible with all experimental constraints. While these states may not be accessible in their direct searches at the LHC in our present scenario, such ewinos
could still be traced in the decays of the top squarks of the above-mentioned kind.
We consider the signal final state 
$ 1\ell \, (e,\mu) + (\geq 1)~ a_s \, (\bbbar) + (\geq 1)~ \hsm \, (\bbbar)+
\geq \, 4 \; \mathrm{jets}  + \etmiss$ at the LHC.
We find that while a 
usual cut-based analysis (CBA) of LHC data worth 300~\fbinv~would be unable to 
discover such excitations, a multivariate analysis (MVA) can be reasonably 
sensitive to higgsino-like ewinos having masses $\gtrsim 650$ GeV 
when $\mstone \gtrsim 1$~TeV. On the other hand, 
with 3000~\fbinv~of data, these masses become accessible in a CBA while even 
an MVA on such a data set is unlikely to find these ewinos with 
masses around 1 TeV when $\mstone$ hits $\sim 1.5$~TeV.
}
\begin{document}
\maketitle
\section{Introduction}
\label{Introduction}
The Next-to-Minimal Supersymmetric Standard Model (NMSSM) was originally motivated as a 
solution to the well-known `$\mu$'-problem~\cite{Kim:1983dt} that plagues the minimal 
supersymmetric (SUSY) extension of the Standard Model (SM), i.e., the MSSM, via the 
introduction of a gauge singlet superfield. The NMSSM is known for long 
to possess a non-trivial phenomenology by virtue of its richer scalar and electroweakino
(ewinos, i.e., charginos and neutralinos) sectors in the presence of singlet
degrees of freedom which can be rather light and have characteristic interactions.

Incidentally, the scenario has attracted a renewed attention post the discovery
of the SM-like Higgs boson ($\hsm$) with a mass ($\mhsm$) around 125 GeV~\cite{ATLAS:2012yve, CMS:2012qbp} at the Large Hadron Collider (LHC). 
This is since while in the MSSM attaining such a mass requires a large radiative
correction and hence much heavier top squarks ($\stonetwo$; running in the
loops) thus rendering the scenario somewhat `unnatural', the NMSSM provides a
more `natural' setup as it finds new tree-level contributions to $\mhsm$ and
hence banks less on too heavy top squarks.

Naturally, various phenomenological aspects of the NMSSM, in particular, of 
its $Z_3$-symmetric incarnation (\z3nmssm), have since been either 
meticulously revisited or have been explored freshly. These not only include 
studies in the usual fronts like the colliders and the dark matter (DM) but, 
of recent, also have opened up to its cosmological implications pertaining to the 
stability of the vacuum~\cite{Beuria:2017wox, Beuria:2016cdk}, phase 
transitions in the early
Universe~\cite{Baum:2020vfl, Athron:2019teq, Chatterjee:2022pxf}, electroweak 
baryogenesis~\cite{Akula:2017yfr}, etc. Such studies have explored the deep 
connections among these fronts, the common thread being the scalar and the 
ewino sectors, now augmented by new gauge-singlet states (two singlet 
scalars and a singlet fermion, i.e., the singlino) when compared to the MSSM 
case.

In this work, a specific choice is made over the neutralino spectrum where the lightest 
SUSY particle (LSP) is bino-like (i.e., $\ntrlone \approx \bino$), the next-to-LSP (NLSP) is singlino-like
(i.e., $\ntrltwo \approx \widetilde{S}$) while the two immediately heavier neutralinos and the lighter chargino ($\ntrlthreefour, \charonepm \approx \widetilde{H}$) are
higgsino-like with very similar masses. 
The wino-like states ($\ntrlfive$ and $\chartwopm$) 
are taken to be much heavier and hence those are effectively decoupled from the phenomenology we  discuss in this work.

Such a hierarchy of ewinos, in particular, with a light LSP (NLSP) which is bino-like (singlino-like), has recently been highlighted~\cite{Abdallah:2020yag} to 
open up the possibility of having a bino-like DM in the \z3nmssm, something that
is highly disfavoured in the MSSM
given  the current experimental constraints from both the collider and the cosmology 
fronts. 
In the \z3nmssm, this is facilitated by a light singlet-like 
$CP$-even scalar, $\hs$, (which necessarily accompanies a relatively light singlino-like state~\cite{Abdallah:2020yag,Ellwanger:2014hia,Ellwanger:2016sur,Ellwanger:2018zxt,Abdallah:2019znp,Guchait:2020wqn,Barman:2020vzm}), and/or a light $CP$-odd scalar, $\as$, 
when these provide efficient funnels for the mutual annihilation of the
bino-like DM, which is tempered with small higgsino and singlino
admixtures.
This is possible without inviting a conflict with various bounds obtained from 
the DM direct detection (DMDD) experiments
(of both spin-independent (SI) and spin-dependent (SD) kinds)
thanks to the occurrence of
new `blind spots'~\cite{Abdallah:2020yag}.
However, in this regard, the tempering of the bino-like LSP with the higgsino admixtures has to, otherwise, maintain a delicate balance. Such a situation, to a good extent, can be ameliorated by pushing $\mueff$
up, optimally, without requiring it to be just that large when the DM annihilation 
rate starts suffering, thus leaving too large a relic.
Such a spectrum of ewinos and light scalar(s) has been
shown~\cite{Abdallah:2020yag} to bear a further theoretical motivation as this can 
trigger a strong first-order electroweak phase transition (SFOEWPT) in the 
early Universe. This could not only set the stage for a successful electroweak baryogenesis 
(EWBG) that might explain the experimentally measured baryon asymmetry in the 
present day Universe, but also could be a potential source of detectable gravitational waves (GWs).

For such a spectrum, direct productions of the NLSP, which is
singlino-dominated, would be naturally suppressed (because of its singlet 
nature) even though it could be reasonably light. Hence this might escape 
searches at the LHC. On the other hand, the higgsino-like states can
stay relatively light  (thus aiding `naturalness') and still remain 
insensitive to searches in their direct productions. This is because their cascades could now involve
singlet-like scalar(s) and the singlino-like neutralino. 

It is to be noted that the most sensitive of searches for relatively light 
ewinos in their direct productions at the 13 TeV LHC involve
multi-lepton plus multi-jet final states accompanied by missing transverse 
energy (MET, $\etmiss$). In particular, the strongest ever lower bounds on 
their masses come from the search in the rather clean $3\ell + \etmiss$ final 
state arising in the direct production of wino-like $\charonepm \ntrltwo$ 
cascading via the $WZ$-mode, where both $W^\pm$- and $Z$-bosons decay 
leptonically~\cite{Sirunyan:2017lae, Aaboud:2018jiw, Aaboud:2018sua, Aad:2019vvi, ATLAS:2020ckz} or from a search in the final state
$2\ell + 2$-${\rm jet} + \etmiss$, where $W^\pm$ decays 
hadronically~\cite{Sirunyan:2017lae, Aaboud:2018jiw, Aaboud:2018sua} or in the 
$WH$-mode~\cite{Sirunyan:2018ubx}, by analyzing the final state 
$1\ell + 2b$-${\rm jet} +\etmiss$, where the $b$-jets originate 
in the decay of~$\hsm$. In contrast, in the \z3nmssm scenario we 
consider in this work, the concerned neutralinos are higgsino-like (instead of 
being wino-like) and the corresponding mass bounds are already expected to be 
relaxed given their smaller combined direct production cross section at the 
LHC. On top of that, as mentioned above, their possible cascades involving 
singlet-like scalars and/or the singlino could further erode their 
sensitivities to current searches at the LHC.

At this point, the theoretical possibility of having relatively light top squarks 
could bring such states into the mix in a rather involved way. As such, 
stringent lower bounds on the mass of the lighter top squark ($\mstone$), as 
obtained by the LHC experiments within the MSSM framework, could still hold in 
an NMSSM scenario as long as $\stone$ decays in ways similar to those in 
the MSSM. However, in the presence of a light singlino-like neutralino and 
light singlet-like scalars, the pattern of cascade decays of $\stone$, via heavier ewinos, could  get altered in an essential way. This could relax the lower bounds on $\mstone$ as reported by the LHC experiments. 
Such modifications can be 
triggered by the altered decay patterns of the somewhat heavier
ewinos, produced in the decays of $\stone$, to their lighter cousins 
which are now accompanied by light singlet-like scalar(s). Note that, as discussed earlier, in the 
first place, such NMSSM-specific cascades of the ewinos could lead to 
relaxed lower bounds on their own masses as obtained by the LHC 
experiments within the MSSM framework.
However, now, in the presence of a relatively light  $\stone$ once the reach for the 
ewinos in their direct (electroweak) productions at the LHC saturates, 
the same can be extended by hunting them down in the cascades of $\stone$ that would be produced strongly at the LHC~\cite{Guchait:2021tmh}.

To be specific, we choose to work with a bino-like LSP with its mass well 
below 100 GeV and a singlino-like NLSP not much heavier than 100 GeV while the 
higgsino-like neutralinos and the higgsino-like lighter chargino have their 
masses in the ballpark of 500 GeV or above, up to about 1 TeV.
Such a choice for the LSP mass, in the presence of an NLSP and
the higgsino-like states of the chosen kinds, is motivated by the fact that these, when acting in tandem, have been demonstrated~\cite{Abdallah:2020yag} to relax the DM direct detection constraints where those tend to get increasingly stronger.
Also, we 
consider $\mstone$ over the range $\sim$ 800~GeV -- 1.5~TeV for our analysis at 
the 13 TeV LHC.\footnote{We carry out our analyses by considering a
13 TeV LHC run that eventually would have collected data worth 3000~\fbinv.
Results obtained with this consideration are unlikely to be affected in any essential way in the upcoming runs of the LHC with a slightly increased centre of mass energy.} 
Note that such spectra are expected to 
evade lower bounds on the involved masses from the LHC experiments, are 
compatible with the constraints from the DM sector and are also able to
trigger SFOEWPT leading to EWBG in the early Universe~\cite{Chatterjee:2022pxf}.
Possible cascades of $\stone$ that we explore in this work (and which lead to our chosen signal final state) are schematically shown in the rectangular box.

As can be seen, for such a spectrum, the ensuing cascades of $\stone$ might
produce $\hsm$ when $\ntrlthreefour$ decay to $\ntrlonetwo$, of which their 
decays to the singlino-like $\ntrltwo$ are of special interest. This is not 
only because a singlet-like scalar, $\as$, accompanies the same but also 
since, over the region of the parameter space that we are interested in, such 
decays of $\ntrlthreefour$ are favored. Hence, as may be expected, the final 
states are going to be rich in bottom quark jets, which will come from the 
decays of $\hsm$, $\as$ and  the top quarks arising from the cascades of $\stone$ 
and/or from $\stone$'s direct decay to a bottom quark with an accompanying lighter 
chargino, $\charonepm$.
Hence the specific signal final state that we study in this work, in the
context of the high luminosity run of the LHC (HL-LHC), is
1 lepton+ 4 $b$-$\mathrm{jets}$ + $\geq$ 4 $\mathrm{jets} + \etmiss$
in which the two pairs of $b$-jets are required to get reconstructed to $\hsm$ 
and $\as$.
\noindent
\begin{figure}[t]
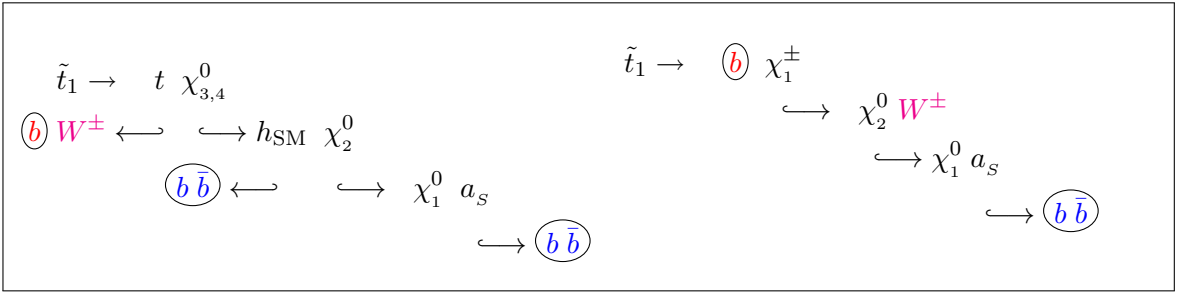

\fbox{
\begin{minipage}{0.455\linewidth}
\begin{equation*}
\begin{aligned}
%
\stone  \to \quad t &
                    \;\;  \ntrlthreefour & & &
                                            \nonumber \\
  \mathcircled{\cred{b}} \;  \cmagenta{W^\pm}  \varlonghookleftarrow & 
                \;\;\; \varlonghookrightarrow 
                                          \hsm
                                    \;\; \ntrltwo &  \nonumber \\
 & \mathcircled{\cblue{b \; \bar{b}}}  \varlonghookleftarrow  \;\;\;\;\;\;\; \varlonghookrightarrow & \ntrlone \;\; & \as \nonumber \\
 & & & \; \varlonghookrightarrow \mathcircled{\cblue{b \; \bar{b}}} \nonumber
 \end{aligned}
 \end{equation*}
 \end{minipage}
\begin{minipage}{.5\linewidth}
\begin{equation*} 
\begin{aligned}
\stone \to \quad \mathcircled{\cred{b}} &  \;\;  \charonepm & & \\
  &  \; \;\; \varlonghookrightarrow & \ntrltwo & \; \cmagenta{W^\pm} \nonumber \\
 & & & \!\!\!\!\! \varlonghookrightarrow \ntrlone \; \as &  \nonumber \\
 & & & & \!\!\!\!\!\!\!\!\! \varlonghookrightarrow \mathcircled{\cblue{b \; \bar{b}}} \nonumber \\
 & & & & \nonumber
\end{aligned}
\label{eq:process}
\end{equation*}
\end{minipage}
}
\caption{Possible cascades of $\tilde{t}_1$ explored in this work that  lead to the chosen signal final state    $1\ell (e, \mu) + (\geq 1) a_s (b  \bar{b}) + (\geq 1) \hsm (b  \bar{b}) + \geq 4 \text{jets} + \etmiss$.}
\end{figure}

A few benchmark points are then picked 
up based on the mass and the dominant chiral content of $\stone$. We perform 
both a cut-based study and a MultiVariate Analysis
(MVA)~\cite{Hocker:2007ht} using the Boosted Decision Tree (BDT)~\cite{Friedman:2002we} and compare their performances in deciphering the signal 
from the background. The sensitivities of the LHC to the proposed signal are further compared for accumulated luminosities of 300~\fbinv~and 3000~\fbinv. 

The paper is organized as follows. In section~\ref{sec:model} we discuss the 
theoretical framework of the \z3nmssm by outlining the scalar (Higgs), the 
ewino and the top squark sectors and highlight how the masses of 
various excitations from these sectors and their mutual interactions, that are 
so crucial for the phenomenology we discuss, depending  on the underlying 
parameters of the theory. Section~\ref{sec:param-space} is devoted to 
establishing 
the region of the \z3nmssm parameter space 
that is particularly interesting for our present purpose. The setup of our 
analysis is presented in section~\ref{sec:setup} which includes discussions of 
various constraints that we consider, both theoretical and experimental, 
the descriptions of the \z3nmssm-specific signal that we are interested in and
the relevant backgrounds as well as the strategy of reconstructing the final 
state objects. Section~\ref{sec:results} discusses in detail the results of 
our simulations by comparing the significances and/or reach of the
\z3nmssm-specific signal that we propose over the SM background at the 13 TeV 
LHC as obtained via a usual cut-based analysis and via a multivariate 
approach. In section~\ref{sec:summary} we conclude.
\section{The theoretical framework: the \z3nmssm} 
\label{sec:model}
The most general \z3nmssm superpotential, with conserved $R$-parity, is given by~\cite{Ellwanger:2009dp}
\beq
{\cal W}= {\cal W}_\mathrm{MSSM}|_{\mu=0} + \lambda \widehat{S}
\widehat{H}_u \cdot \widehat{H}_d
        + {\kappa \over 3} \widehat{S}^3 \, ,
\label{eqn:superpot}
\eeq
where ${\cal W}_\mathrm{MSSM}|_{\mu=0}$ represents the MSSM superpotential without the higgsino mass term (known as the $\mu$-term), $\widehat{H}_u, \widehat{H}_d$ and $\widehat{S}$ are the $SU(2)$ Higgs doublet superfields and the gauge
singlet superfield, respectively, and `$\lambda$' and `$\kappa$' are dimensionless coupling constants. The addition of a singlet superfield $\widehat{S}$ to the MSSM helps solve the so-called $\mu$-problem when its scalar component develops a non-zero vacuum expectation value~(\vev)~$\vs$ thereby generating an effective $\mu$-term dynamically which is given by $\mueff=\lambda \vs$ and can be obtained from the second term on the right-hand side of equation~\ref{eqn:superpot}. The corresponding soft SUSY-breaking Lagrangian is given by
\beq
-\mathcal{L}^{\rm (soft)}= -\mathcal{L_{\rm MSSM}^{\rm (soft)}}|_{B\mu=0}+ m_{S}^2
|S|^2 + (
\lambda A_{\lambda} S H_u\cdot H_d
+ \frac{\kappa}{3}  A_{\kappa} S^3 + {\rm h.c.}) \,,
\label{eqn:lagrangian}
\eeq
where `$B$' is the soft SUSY-breaking bilinear parameter of the MSSM associated with the Higgsino mass term, $m_S$ is the soft SUSY-breaking mass
of the singlet scalar field `$S$' while $\alambda$ and $\akappa$ are the NMSSM-specific trilinear soft couplings, all of which are of mass dimension one. In this study, we do not consider $CP$-violating entries that might be present in the scalar and/or the ewino sectors. Hence all the 
Lagrangian parameters are taken to be real. In the upcoming subsections, we briefly discuss the structures of the Higgs, the ewino and the top squark sectors and the  interactions involving them which are instrumental in the phenomenology we discuss in the rest of this work.
%
\subsection{The Higgs sector}
\label{subsec:higgs-sector}
The \z3nmssm has the following tree-level Higgs (scalar) potential: 
%
\begin{subequations}
\begin{eqnarray}
V_{\textrm{tree}}^{\textrm{NMSSM}} = V_F + V_D + V_\textrm{soft} \, ,
\label{Eq:NMSSMHiggspotential}
\end{eqnarray}
%
where contributions from the $F$- and $D$-terms as well as those from the soft SUSY--breaking terms, as represented by the variables $V_F$, $V_D$ and $V_\textrm{soft}$, respectively, are given by
\%begin{align}
\begin{eqnarray}
V_F &=& \left |\lambda S \right|^2 (|H_u|^2 + |H_d|^2) + \left |\lambda H_u \cdot
H_d + \kappa S^2 \right |^2 , \\
V_D &=& \frac{1}{8} g^2(|H_u|^2 - |H_d|^2)^2 + \frac{1}{2} g_2^2|H_u^\dagger H_d|^2 \,,\\
V_{\textrm{soft}} &=& m_{H_u}^2 |H_u|^2 + m_{H_d}^2 |H_d|^2 + m_{S}^2
|S|^2 + (\lambda A_\lambda S H_u \cdot H_d +  \frac{\kappa}{3} A_\kappa S^3 +
  \textrm{h.c.}) \, ,
\end{eqnarray}
\end{subequations}
where $g^2 = (g_1^2 + g_2^2)/2$ and $g_1$ and $g_2$ are the $U(1)$ and the $SU(2)$ gauge 
couplings, respectively.  The complex Higgs fields can be then 
expressed as 
\begin{align}
H_u= \begin{pmatrix} H_u^+\\ v_u + \tfrac{1}{\sqrt{2}} \left(H_{uR} + iH_{uI}\right) 
\end{pmatrix}, \quad
H_d= \begin{pmatrix} v_d + \tfrac{1}{\sqrt{2}} \left(H_{dR} + iH_{dI}\right) \\ H_d^- 
\end{pmatrix}, \quad
S = \vs + \frac{S_R + iS_I}{\sqrt{2}},
\label{complexfields}
\end{align}
where suffixes `$R$' and `$I$' stand for the $CP$-even and the $CP$-odd components of the complex fields and 
 $v_u$, $v_d$ and $v_s$ are the \vevs~ 
 of the $CP$-even components of the neutral scalar fields with $\sqrt{\vu^2 + \vd^2}= v \simeq 174$ 
GeV, $\tanb= \vu/\vd$ and $\mueff= \lambda \vs$ (as mentioned earlier).
Thus, the symmetric (3$\times$3) mass-squared matrix for the $CP$-even Higgs bosons of the $Z_3$-symmetric NMSSM, in the basis
$H_{jR}=\{H_{dR}, H_{uR}, S_R\}$, is given by~\cite{Ellwanger:2009dp}
\beq
{\small{
{\cal M}_S^2 =
\left( \begin{array}{ccc}
  g^2 \vd^2 + \mueff \beff \,\tanb
& \:\; (2\lambda^2 - g^2) \vu \vd - \mueff \beff
& \:\; \lambda (2 \mueff \, \vd - (\beff + \kappa \vs) \vu) \\[0.2cm]
 \ldots
& \:\; g^2 \vu^2 + \mueff \beff /\tanb
& \:\; \lambda (2 \mueff \, \vu - (\beff + \kappa \vs)v_d) \\[0.2cm]
 \ldots
 &  \ldots
& \:\:\, \lambda \alambda  \frac{\vu \vd}{\vs} + \kappa \vs (\akappa + 4\kappa \vs)
\label{eqn:cp-even-matrix} 
\end{array} \right),}
}
\eeq
where $B_{\mathrm{eff}}=\alambda+\kappa\vs$ is the effective `$B$' term of the \z3nmssm.
The resulting $CP$-even (scalar) mass eigenstates are given by
\bea
h_i &=& S_{ij} H_{jR}, \qquad  
\mathrm{with} \quad {i,j=1,2,3}\, ,
\label{eqn:cp-even-scalar-physical-states}
\eea
where the matrix `$S$' diagonalizes ${\cal M}_{S}^2$.
In a rotated basis ($\hat{h}, \widehat{H}, \hat{s}$)~\cite{Miller:2003ay, Badziak:2015exr}, where $\hat{h} = H_{dR} \cos{\beta} + H_{uR} \sin{\beta}$, $\widehat{H} = H_{dR} \sin{\beta} - H_{uR} \cos{\beta}$ and $\hat{s} = S_{R}$,
the $CP$-even state $\hat{h} \, (\widehat{H})$ now mimics the SM Higgs (MSSM (doublet)-like heavier $CP$-even Higgs boson, `$H$') field. In this basis, the $CP$-even scalar mass eigenstates posses the following generic admixtures: 
\bea
h_i &=& E_{h_i \hat{h}} \hat{h} + E_{h_i \widehat{H}} \widehat{H} + E_{h_i \hat{s}} \hat{s} \,,
\label{eqn:hiEhhat}
\eea
where the matrix $E_{ab}$ diagonalizes the mass-squared
matrix for the $CP$-even scalars in the rotated basis. On the other hand, the symmetric (3$\times$3) mass-squared matrix for the $CP$-odd scalars in the basis
$H_{jI}=\{H_{dI}, H_{uI}, S_I\}$ is given by~\cite{Ellwanger:2009dp}
\beq
{\small{
{\cal M'}_{P}^2 =
\left( \begin{array}{ccc}
  \mueff \beff \,\tanb
& \mueff \beff
& \lambda \vu (\alambda - 2\kappa \vs) \\[0.2cm]
  \ldots
& \mueff \beff /\tanb
& \lambda \vd (\alambda - 2\kappa \vs) \\[0.2cm]
  \ldots
&  \ldots
& \lambda (\beff + 3 \kappa \vs) \frac{\vu \vd}{\vs} - 3 \kappa \akappa \vs
\label{eqn:cp-odd-matrixp}
\end{array} \right).}
}
\eeq
For ${\cal M'}_{P}^2$,  a similar rotation in the basis of the doublet states $H_{dI}$ and $H_{uI}$ would, as can be expected, projects out the massless Nambu-Goldstone mode which can be dropped. The (doublet)-like
$CP$-odd (pseudoscalar) Higgs boson of the MSSM can be identified with $A=\cos\beta H_{uI} + \sin\beta H_{dI}$  and in the basis $\{A, S_I \}$, the symmetric (2$\times$2) mass-squared matrix for
the $CP$-odd scalars of the $Z_3$-NMSSM gets reduced to
\beq\label{eqn:cp-odd-matrix}
{\small
{\cal M}_P^2 =
\left( \begin{array}{cc}
    m_A^2
~&~ \lambda (\alambda - 2\kappa \vs)\, v \\[0.2cm]
    \lambda (\alambda - 2\kappa \vs)\, v
~&~ \lambda (\alambda + 4\kappa \vs)\frac{v_u v_d}{\vs} -3\kappa \akappa \, \vs  
\end{array} \right),
}
\eeq
where, along the lines of the MSSM, $m_A^2= 2 \mueff \beff / \sin2\beta$. The $CP$-odd (pseudoscalar $a_k$) mass eigenstates are then given by
\bea
a_k = {\cal O}_{kA} A + {\cal O}_{kS_I} S_I, 
\qquad \mathrm{with} \quad {k=1,2},
\label{eqn:cp-odd-scalar-physical-states}
\eea
where the matrix `${\cal O}$' diagonalizes ${\cal M}_{P}^2$.

For a negligible singlet-doublet ($\{H_{dR}$-$H_{uR}\}$-$S_R$) mixing, the squared mass (at the tree-level) of the 
singlet-like $CP$-even physical state can be approximated as
\beq
\mhssq \approx {\cal{M}}^2_{S,33} =  
 \lambda \alambda  \frac{v_u v_d}{\vs} + \kappa \vs (\akappa + 4\kappa \vs) \,.
\label{eqn:cp-even-mass}
\eeq
For $m_A >> m_Z$ (i.e., in the so-called decoupling limit), the masses of the heavier of the MSSM (doublet)-like $CP$-even Higgs boson (`$H$') and the charged Higgs bosons ($H^{\pm}$) approach $m_A$, i.e., $m_H \approx m_{H^{\pm}} \approx m_A$. On the other hand, the squared mass of the
lighter one mimicking $\hsm$ is given by
\cite{Ellwanger:2011sk}
\beq
\mhsm^2 = m_Z^2 \cos^2 2\beta + \lambda^2 v^2 \sin^2 2\beta + \Delta_\mathrm{mix}+ 
\Delta_{\mathrm{rad.\,corrs.}} \, ,
\label{eqn:hsmmass}
\eeq  
where the first term on the right-hand side is the tree-level MSSM contribution, while the second and the third terms are new tree-level contributions from the NMSSM. $\Delta_\mathrm{mix}$ has its origin in a possible singlet-doublet mixing and in the limit of a weak mixing, $\Delta_{\rm mix}$ is given by~\cite{Ellwanger:2011sk}
\beq
\label{eqn:hmix}
\Delta_{\rm mix} = \frac{4 \lambda^2 \vs^2 v^2 
(\lambda -\kappa \sin 2\beta)^2} {\tilde{m}_h^2-m_{ss}^2} \, ,
\eeq
where 
$m_{ss}^2 = \kappa \vs(A_{\kappa}+4 \kappa \vs)$ and $\tilde{m}_h^2 = \mhsm^2 - \Delta_{\rm mix}$. For $\vs >> \vu, \vd$, one finds $m_{ss}^2 \approx {\cal{M}}^2_{S,33} \approx \mhssq$.
The term $\Delta_{\mathrm{rad.\,corrs.}}$ captures the dominant radiative corrections in the MSSM at one-loop level (with the top quark and the top squarks running in the loops) to
$\mhsm^2$ and is given by
\cite{Haber:1996fp, Djouadi:2005gj}
\beq
\Delta_{\mathrm{rad.\,corrs.}}^{{\rm 1-loop}}\simeq \frac{3 m_t^4}{4 \pi^2 v^2 \sin^2 \beta}
\left[2 \log \frac{M_S}{m_t}
+ \frac{X_t^2}{ M_S^2} \left(1- \frac{X_t^2}{12 M_S^2} \right) \right],
\label{eqn:radcorr}
\eeq
where $m_t$ denotes the mass of the SM top quark,
$M_S=\sqrt{m_{\widetilde{t}_1} m_{\widetilde{t}_2}}$,
$m_{\widetilde{t}_1 (\widetilde{t}_2)}$ being the mass of the lighter (heavier) 
physical top squark state and $X_t=A_t-\mueff \cot\beta$, $A_t$ being the soft 
trilinear coupling for the top sector.

As for the singlet $CP$-odd scalar, $\as$, its squared mass, up to a small 
mixing with the doublet-like $CP$-odd state, is given by
\beq
\massq \approx {\cal{M}}^2_{P,22} =  
\lambda (\alambda + 4\kappa \vs)\frac{v_u v_d}{\vs} -3\kappa \akappa \, \vs \, .
\label{eqn:cp-odd-mass}
\eeq
The expressions for $\mhsm, m_{H,A}, \mhs$ and $\mas$ presented above reveal their
nontrivial dependencies on various input parameters of the $Z_3$-NMSSM and point to characteristic correlations among these masses.
Together, these could crucially
govern the collective phenomenology involving these states at both colliders and at the DM/cosmology front.
%
\subsection{The ewino sector}
\label{subsec:ewino-sector}
The neutralino sector of the \z3nmssm consists of five neutralinos
which are mixtures of the bino ($\bino$), the wino ($\wino^0_3$), two higgsinos 
($\higgsinod$, $\higgsinou$) and  the singlino ($\singlino$). The real symmetric 
($5\times 5$) neutralino mass-matrix, ${\cal M}_0$, in the basis
$\psi_0 \equiv \{\widetilde{B},~\widetilde{W}^0_3, ~\widetilde{H}_d^0,~\widetilde{H}_u^0, ~\widetilde{S}\}$, is given by~\cite{Ellwanger:2009dp}
\beq
\label{eqn:mneut}
{\cal M}_0 =
\left( \begin{array}{ccccc}
\mone & 0 & -\dfrac{g_1 \vd}{\sqrt{2}} & \dfrac{g_1 \vu}{\sqrt{2}} & 0 \\[0.4cm]
\ldots & \mtwo & \dfrac{g_2 \vd}{\sqrt{2}} & -\dfrac{g_2 \vu}{\sqrt{2}} & 0 \\
\ldots & \ldots & 0 & -\mueff & -\lambda \vu \\
\ldots & \ldots & \ldots & 0 & -\lambda \vd \\
\ldots & \ldots & \ldots & \ldots & 2 \kappa \vs
\end{array} \right) \,,
\eeq
where $M_{1,2}$ are the soft SUSY-breaking masses for the $U_1$ and the $SU(2)$ 
gauginos, i.e., the bino and the wino, respectively. The [5,5] element of
${\cal M}_0$, i.e., $2\kappa \vs \equiv \msinglino$, is the singlino mass term. 
The matrix ${\cal M}_0$ can be diagonalized by an orthogonal $5 \times 5$ matrix, 
`$N$', i.e.,
\beq
N {\cal M}_0 N^T= {\cal M}_D= {\rm diag}(m_{{_{\chi}}_{_1}^0},m_{{_{\chi}}_{_2}^0},m_{{_{\chi}}_{_3}^0},m_{{_{\chi}}_{_4}^0},m_{{_{\chi}}_{_5}^0})  \, , 
\label{eqn:diagonalise-1}
\eeq
with growing $\mntrli$ as `$i$' increases whereas the neutralino mass eigenstates,
$\ntrli$, are given by
\beq
\ntrli = N_{ij} \psi_j^0 \, , \quad \text{with} \;\; i,j=1,2,\ldots,5 \, .
\label{eqn:diagN2}
\eeq
As we set $\mtwo$ large, the heaviest neutralino ($\ntrlfive$) is a nearly pure 
wino with $\mntrlfive \approx \mtwo$ and is practically decoupled for our purposes.
The lightest neutralino ($\ntrlone$) can get to be the lightest of the SUSY particles and, while $R$-parity is conserved, such an $R$-parity odd state becomes stable and could turn out to be a potential candidate for the cold DM.

The chargino sectors of the  MSSM and that of the \z3nmssm are structurally 
identical but for $\mu$ $\rightarrow$ $\mueff$ for the latter. Thus, the
$2 \times 2$ chargino mass-matrix, ${\cal M}_C$, in the gauge bases
$\psi^+ = \{ -i \widetilde{W}^+, \, \widetilde{H}_u^+ \}$ and
$\psi^- = \{ -i \widetilde{W}^-, \, \widetilde{H}_d^- \}$,
is given by~\cite{Ellwanger:2009dp}
\beq
{\cal M}_C = \left( \begin{array}{cc}
                    \mtwo   & \quad  g_2 \vu \\
                 g_2 \vd  & \quad \mueff 
             \end{array} \right) .
\eeq
It requires two $2 \times 2$ unitary matrices `$U$' and `$V$' to diagonalize this 
asymmetric matrix, ${\cal M}_C$~\cite{Gunion:1984yn}, i.e.,
\beq
U^* {\cal M}_C V^\dagger = \mathrm{diag} (\mcharone , \mchartwo) \; , \quad
\mathrm{with} \;\;  \mcharone < \mchartwo  \; .
\label{eqn:uvmatrix}
\eeq
With $\mtwo$ being set large in this work, $\charonepm$ is higgsino-like while the
wino-like $\chartwopm$ is essentially decoupled from the phenomenology we discuss.
\subsection{The top and bottom squark sector}
\label{subsec:stop}
%
The tree-level mass-squared matrices for the top and bottom squarks in the NMSSM, in the basis 
of the weak eigenstates $\{\stleft, \stright\}$ and $\{\sbleft, \sbright\}$, is given by, respectively,
\beq\label{eqn:mstop} 
{\cal M}_{_{\tilde{t}}}^2= \left(\ba{cc}  m_{\stleft}^2
& m_t X_t^{\dagger}  \\
 m_t X_t & 
m_{\stright}^2
\ea \right),
\eeq
and
\beq\label{msbottom}
{\cal M}_{_{\tilde{b}}}^2= \left(\ba{cc}  m_{\sbleft}^2
& m_b X_b^{\dagger}  \\
 m_b X_b & 
m_{\sbright}^2
\ea \right),
\eeq
where
$m_{\stleft}^2 = m_{\tilde{Q}_{3L}}^2 + y_t^2 v_u^2 +
 (v_u^2-v_d^2)\left(\frac{g_1^2}{12}-\frac{g_2^2}{4}\right)$,
$m_{\stright}^2 = m_{\tilde{U}_{3R}}^2 + y_t^2 v_u^2-(v_u^2-v_d^2)\frac{g_1^2}{3}$,
$m_{\sbleft}^2 = m_{\tilde{Q}_{3L}}^2 + y_b^2 v_d^2 + (v_u^2-v_d^2)\left( \frac{g_1^2}{12}+\frac{g_2^2}{4}\right)$,
 $m_{\sbright}^2 = m_{\tilde{D}_{3R}}^2 + y_b^2 v_d^2 +(v_u^2-v_d^2)\frac{g_1^2}{6}$
with $m_{\widetilde{Q}_{3L}}$, $m_{\widetilde{U}_{3R}}$ and $m_{\widetilde{D}_{3R}}$ being the soft masses for the $SU(2)$ doublet, the up-type and the down-type $SU(2)$ singlet squarks from the third generation, respectively.
The squark mass eigenstates, $\tilde{f}_1$ and 
$\tilde{f}_2$ ($m_{\tilde{f}_1} < m_{\tilde{f}_2}$) where $\tilde{f}$ stands for $\tilde{t}$ or $\tilde{b}$, respectively, are found on diagonalizing the matrix
${\cal M}_{_{\tilde{f}}}^2$ of equation (\ref{eqn:mstop}) using a unitary matrix
${\cal R}_{\tilde{f}}$ given by
\beq
\label{stopmixmatrix}
 {\cal R}_{\tilde{f}} = \left( 
                     \begin{array}{rr}
                       \cos\theta_{\tilde{f}} & \: \sin\theta_{\tilde{f}} \\
                      -\sin\theta_{\tilde{f}} & \: \cos\theta_{\tilde{f}}
                     \end{array}
              \right) \; ,
\eeq
where $\theta_{\tilde{f}}$ is the squark mixing angle given by
\beq 
\sin 2\theta_{\tilde{f}} = {2m_f X_f \over {m_{\tilde{f}_2}^2 - m_{\tilde{f}_1}^2}} \quad, \qquad X_f =A_{\tilde{f}} -\mueff \; r \quad ,
\; 
\eeq
where $r= \cot\beta \, (\tan\beta)$ for the stop (sbottom) sector.
Note that, in this formalism, $\tilde{f}_1 \equiv \tilde{f}_L (\tilde{f}_R)$ when $\theta_{\tilde{f}}=0 ({\pi \over 2})$.
One could expect a light
bottom squark along with a light top squark close by in mass when $m_{\tilde{Q}_3}$ is relatively small and both of them being dominantly left-handed type.
%
\subsection{Interactions of the ewinos with $\stone$, the Higgs and the gauge bosons}
\label{subsec:interactions}
We are now all set to discuss briefly the interactions of various 
ewinos with $\stone$ and with the pertinent Higgs and gauge bosons 
which determine the nature and the strengths of the cascades that $\stone$ 
undergoes. 
\subsubsection{Interactions of the ewinos with $\stone$}
\label{subsubsec:ewino-stop1}
Decays of pair-produced $\stone$-s to the ewinos
($\ntrli \, (i \neq 5)$ and $\charonepm$) that are central to the present work 
are governed by the following Lagrangian pieces:
\begin{subequations}
\begin{eqnarray}
\mathcal{L}_{\stone \bbar \, \charonep} &=& \bbar \left( f^{\charonep}_L P_L + f^{\charonep}_R P_R \right) {{\charonep}^c} \, \stone  +\textrm{h.c.} \, ,
\label{eqn:bt1chi} \\
\mathcal{L}_{\stone \tbar \, \ntrli} &=& \tbar \left( f^{\ntrli}_L P_L + f^{\ntrli}_R P_R \right) {\ntrli} \, \stone  +\textrm{h.c.} \, ,
\label{eqn:tt1chi}
\end{eqnarray}
\label{eqn:stop-ewino}%
\end{subequations}
where $P_{L,R}=\frac{1}{2}\left(1\mp\gamma^{5}\right)$ are the familiar projection 
operators and the various coefficients appearing in these equations are given by
\begin{subequations}
\begin{eqnarray}
f^{\charonep}_{L}&=&y_{b} \, U^{*}_{12}\cos\theta_{\tilde{t}} \; , \label{eqn:fcharginoL} \\
f^{\charonep}_{R}&=&-g_{2}V_{11}\cos\theta_{\tilde{t}}+y_{t}V_{12}\sin\theta_{\tilde{t}} \; , \label{eqn:fcharginoR}
\end{eqnarray}
\label{eqn:stop-chargino}
\end{subequations}
\vskip -35pt
\begin{subequations}
\begin{eqnarray}    
f^{\ntrli}_{L}&=&-\left[\frac{g_{1}}{3\sqrt{2}}N_{i1} + \frac{g_{2}}{\sqrt{2}}N_{i2} \right]\cos\theta_{\tilde{t}}-y_{t}N_{i4}\sin\theta_{\tilde{t}} \; , \label{eqn:fchiL} \\
f^{\ntrli}_{R}&=&\frac{2\sqrt{2}}{3}g_{1}N^{*}_{i1}\sin\theta_{\tilde{t}}-y_{t}N^{*}_{i4}\cos\theta_{\tilde{t}} \; ,
\label{eqn:fchiR}
\end{eqnarray}
\label{eqn:stop-neutralino}%
\end{subequations}
where $y_{b(t)}=\sqrt{2}m_{b(t)}/v\cos(\sin)\beta$ is the bottom (top) quark 
Yukawa coupling. Note that as we set $\mtwo$ large,
$V_{11}, U_{11}, N_{i2} \; (i \neq 5)$ are minuscule and can safely be ignored
in this work.

With these in mind, a brief discussion on the implications of equations
\ref{eqn:stop-ewino}, \ref{eqn:stop-chargino} and \ref{eqn:stop-neutralino} is 
in order. The way $\thetastop$ is defined (see above), interactions of
$\stone \sim \stleft \, (\stright)$ with the ewinos are going to be
$\costhetastop \, (\sinthetastop)$-enhanced. Thus, for $\stone \sim \stleft$,
the term $y_{t}N^{*}_{i4}\cos\theta_{\tilde{t}}$, driven by $y_t$, in equation 
\ref{eqn:fchiR}, would clearly dominate over the term driven by $y_b$ in 
equation~\ref{eqn:fcharginoL} (for sure, for low to moderate values of 
$\tanb$) and over the first term in equation \ref{eqn:fchiL} while all other 
contributions could be ignored. Phenomenologically, this is expected to make the 
decay $\stone (\sim \stleft) \to \ntrlthreefour$ dominate over its decays to
$b \charonepm$ and $t \ntrlonetwo$. On the other hand, for
$\stone (\sim \stright)$, all three $\sinthetastop$-dependent terms in equations
\ref{eqn:fcharginoR}, \ref{eqn:fchiL} and \ref{eqn:fchiR} might get comparable.
Hence decays of $\stone$ to all related modes could compete, albeit the one 
to $b \charonepm$ might enjoy some edge due to a larger phase space available to 
it. These considerations are going to be crucial in our analysis and for what we 
present subsequently in section~\ref{sec:results}. 
%

\subsubsection{Interactions of the ewinos with the Higgs and the gauge bosons}
\label{subsubsec:ewino-higgs-gauge}
The pertinent interactions in the current 
category are comprised of the ones like $\ntrli$--$\ntrlj$--$\Phi/Z$
(with $i,j \in \{1,2,3,4\}$) where $\Phi$ includes the neutral Higgs bosons of both 
$CP$-even ($h_i$, including $\hsm$) and $CP$-odd ($a_i$) kinds and the ones involving $\charonepm$--$\ntrlonetwo$--$W^\pm$.

Given that the interactions involving a pair of neutralinos are crucial for the
collider and the DM phenomenologies alike, we present those below in their 
general form~\cite{Abdallah:2020yag}. Following the convention introduced in 
section \ref{subsec:higgs-sector}, the generic  such interaction involving a 
$CP$-even scalar, $h_i$, in the basis presented in equation~(\ref{eqn:hiEhhat}), 
is given by
\bea
\label{eqn:hinjnk}
g_{_{h_i \ntrlj \ntrlk}}
 & = &
\Bigg[{\lambda \over \sqrt{2} } \big[ E_{h_i \hat{h}} N_{j5} (N_{k3} \sin\beta + N_{k4} \cos\beta)
 + E_{h_i \hat{H}} N_{j5} (N_{k4} \sin\beta - N_{k3} \cos\beta)  \nonumber \\
 & & \hskip 25pt + \; E_{h_i \hat{s}} (N_{j3}N_{k4} - {\kappa \over \lambda} N_{j5}N_{k5})\big] + {1 \over 2}\big[g_{1}N_{j1} - g_{2}N_{j2}\big] \big[E_{h_i \hat{h}} (N_{k3} \cos\beta - N_{k4} \sin\beta) \nonumber \\
 & & \hskip 25pt + E_{h_i \hat{H}} (N_{k3} \sin\beta + N_{k4} \cos\beta)\big]\Bigg] + \Bigg[j \longleftrightarrow k \Bigg]\, ,
\eea
while the same involving a $CP$-odd scalar, $a_i$, in the basis shown in
equation~(\ref{eqn:cp-odd-matrix}), reads
\bea
\label{eqn:ainjnk}
g_{_{a_i \ntrlj \ntrlk}}
 & = &
\Bigg[i\Big({\lambda \over \sqrt{2}} \big[{\cal O}_{iA} N_{j5} (N_{k4} \sin\beta + N_{k3} \cos\beta) + {\cal O}_{iS_I} (N_{j3}N_{k4} - {\kappa \over \lambda} N_{j5}N_{k5})\big]  \nonumber \\
 & & \hskip 25pt - \; {1 \over 2}\big[g_{1}N_{j1} - g_{2}N_{j2}\big] \big[{\cal O}_{iA} (N_{k3} \sin\beta - N_{k4} \cos\beta)\big]\Big)\Bigg] + \Bigg[j \longleftrightarrow k \Bigg] . \quad
\eea
Note that when $\mtwo$ is taken to be very large, only $j,k \in \{1,2,3,4\}$ are relevant and $N_{j2,k2} \sim 0$.

On the other hand, the interactions of the $Z$-boson with a pair of 
neutralinos and those of the $W^\pm$-boson with a neutralino and a chargino 
are given, respectively, by
\bea
g_{_{Z \ntrli \ntrlj}}
  &=& 
 {g_2 \over {2 \cos\theta_W}} \left( N_{i3} N_{j3} - N_{i4} N_{j4} \right), 
 \label{eqn:zninj} \\
 & & \nonumber \\
g_{_{W^\pm \ntrli \charjmp}}
  &\sim &
  {g_2 \over \sqrt{2}} \left( N_{i4} V_{j2} P_L - N_{i3} U_{j2} P_R \right)
  \, ,
\label{eqn:wnicj}
\eea  
with $U_{j1},V_{j1} \sim 0$ as $\mtwo$ goes large.

For both the scalar and the pseudoscalar interactions with the ewinos, two 
general observations can be made from equations
\ref{eqn:hinjnk} and
\ref{eqn:ainjnk} which are much relevant in the 
context of the present work.
First, neutral doublet scalars interact with a pair of neutralinos via their 
higgsino and singlino contents the strengths of which go as `$\lambda$'. In 
contrast, their singlet counterparts interact with a pair of neutralinos 
exclusively via the higgsino or the singlino contents in the latter. While the 
first type of interactions are driven by `$\lambda$', the latter is 
proportional to `$\kappa$'.
Second, such interactions that are driven by the gauge couplings (in 
particular, $g_1$, which only is relevant for the present work as we set winos 
to be rather massive) always look for the gaugino and the higgsino contents in 
the two neutralinos and involve only the doublet scalars.

As for the interactions of the $Z$-boson to a pair of neutralinos, 
equation \ref{eqn:zninj} indicates that these depend only on the 
higgsino contents of the involved neutralinos. Similarly, the same 
among the $W^\pm$-boson and a neutralino and a chargino depend on the 
higgsino contents of the latter two (albeit only when the wino 
is set very heavy) as can be seen in equation~\ref{eqn:wnicj}.

Further, it is important to note that, in regard to decays of a heavier 
ewino to a much lighter one, both neutralinos can have reasonable 
higgsino contents only via mixings of higgsinos with other 
gaugino/singlino states. In the present work, we bank on such mixings
with the singlino, which is again driven by `$\lambda$' in the \z3nmssm.
Hence decays of higgsino-dominated ewinos to a
singlino-dominated one, either with an accompanying scalar (Higgs) or a 
gauge boson, would be enhanced for larger values of `$\lambda$' when these decay modes are expected to compete as well.
%
\section{The motivated region of the \z3nmssm parameter space}
\label{sec:param-space}
The region of \z3nmssm parameter space that we target for the present study is
guided by several considerations pertaining to the phenomenology of the scenario at
the LHC and that relates to the DM sector. These include the requirements of a 
reasonably light $\stone$, which has a substantial cross section for being
pair-produced at the LHC, something that we exploit directly in this work and
a spectrum of relatively light ewinos with a bino-dominated LSP which
can be a viable DM candidate only when accompanied by not so heavy
singlino- and higgsino-like states and light enough singlet-like scalar(s),
all of which might escape searches at the LHC till date \cite{Abdallah:2020yag}. 
Moreover, as pointed out earlier, such a spectra of ewinos and
singlet scalars can be further motivated~\cite{Abdallah:2020yag} on cosmological 
grounds as these could facilitate a strong first-order electroweak phase
transition (SFOEWPT) in the early Universe, thus paving the way for a successful 
electroweak baryogenesis (EWBG). However, given that in this work, we are mainly interested in the signatures at the LHC, we do not carry out a dedicated analysis of the nature of the phase transitions.

It is interesting to note that the chance of having a light
bino-dominated LSP (with its mass around tens of GeV) that is a viable DM 
candidate is enhanced in the \z3nmssm~\cite{Abdallah:2020yag} when the same is 
moderately tempered with singlino and higgsino admixtures. This certainly 
requires the relevant mass-parameters, i.e., $\mone$, $\msinglino$ and 
$\mueff$, to lie close by and hence the physical masses of these states to be 
of comparable magnitudes. Even then, an optimal tempering is possible only for 
larger values of `$\lambda$' for which the required rate of DM annihilation is 
ensured such that it does not over-close the Universe.

However, such a tempering, in turn, would tend to enhance the DMDD spin-independent (DMDD-SI)  
rate to a level that is ruled out by experiments~\cite{Baum:2017enm}. Interestingly enough, 
it has recently been pointed out~\cite{Abdallah:2020yag} that an occurrence of a 
so-called coupling `blind spot' ($g_{_{\hsm \ntrlone \ntrlone}} \sim 0$) could 
save the situation. For the kind of spectrum of ewinos (i.e.,
{\footnotesize{$\mntrlone{_{(\bino)}} < \mntrltwo{_{(\singlino)}} < \mntrlthreefour{_{(\widetilde{H})}} \approx \mcharone{_{(\tilde{H})}}$}})
that we consider, this is realized generically if
\bea
\label{eqn:blindspot}
\bigg(\mntrlone + {2 \lambda^2 v^2 \over {\msinglino - \mntrlone}}\bigg) {1 \over \mueff \sin 2\beta} \simeq -1 \,.
\eea
Note that for a vanishing `$\lambda$', one retrieves the well-known blind spot condition in the MSSM, i.e., 
{\small $\mntrlone / \mueff \approx -\sin 2\beta \, ,$}
which implies $\mntrlone$ and $\mueff$ to have a relative sign between them.
A closer inspection of equation \ref{eqn:blindspot} reveals that for a finite 
`$\lambda$', as is the case in the NMSSM, there is a further possibility of 
realizing a blind spot even when $\mntrlone$ and $\mueff$ carry the same sign, 
only if there is a relative sign between $\mueff$ and $\msinglino$ (or 
`$\kappa$', for that matter).
Incidentally, again a moderate to the large value of `$\lambda$' is 
required for the blind spot mechanism to work.

These prompt us to work with larger values of `$\lambda$'.
Such a choice has an immediate implication for 
$\mhsm$ as the latter could now be found in the right ballpark even when  
$\stone$ is reasonably
light~\cite{Ellis:1986yg, Barbieri:1987fn, Baer:2012up,Baer:2012cf} which, in 
turn, renders the scenario more `natural' and hence, theoretically more 
attractive. To our advantage, this would boost the production of a 
$\stone$-pair at the LHC. In passing, we note that (as is well known) the chiral contents of 
$\stone$ may also crucially determine its decays (see 
section~\ref{subsubsec:ewino-stop1}), thus affecting the sensitivities in its 
search. We take this issue up in section~\ref{subsec:benchmarks}.

Thus, we find that the multiple roles played by a relatively 
large `$\lambda$' could all work in tandem to our advantage. 
Note, further, that optimally low values of `$\kappa$' and $\mueff$ are 
required to find the singlino-dominated NLSP close by in mass to the LSP to 
facilitate an optimal tempering of an otherwise bino-dominated LSP. 
Furthermore, the smaller the value of $\mueff$ that we could arrange to accommodate in our scenario (which evades 
the stringent LHC constraints), the more attractive would be the scenario on the `naturalness' ground.

In this work, a relatively large `$\lambda$' continues to play a favorable role 
in the collider domain as well. Given that for an enhanced `$\lambda$', the 
mixing among the higgsino and the singlino states gets to be appreciable, 
decays of the higgsino-dominated states to a singlino-dominated 
state are preferred over those to the bino-dominated LSP, i.e.,
{\footnotesize $\mathrm{BR}\left[\widetilde{H} \to \widetilde{S} \; \hsm/Z/W^{\pm} \right] >  \mathrm{BR}\left[ \widetilde{H} \to \widetilde{B} \; \hsm/Z/W^{\pm} \right]$.} 
Considerations of the altered decay patterns of the higgsino-like states, 
$\ntrlthreefour$, that the presence of such an intermediate $\ntrltwo$ triggers, would inevitably relax the stringent lower bounds on the masses of the 
former that are obtained by the LHC experiments by analyzing the final states, 
$3\ell + \etmiss$~\cite{Sirunyan:2018ubx, Aad:2019vvi, ATLAS:2021moa, CMS:2021cox} and $1\ell+2b$-jets+$\etmiss$~\cite{Aad:2019vvf, CMS:2017kyj}. Along the line with what has been shown in an MSSM setup in reference~\cite{Guchait:2021tmh},  
here also some strengthening of the lower bounds on the masses of the ewinos, which previously got relaxed in their direct 
searches when assuming their altered decay patterns, is envisaged when searched for in the cascades of top squark pairs.
Moreover, given that the altered cascade patterns  of the ewinos imply the same for 
the cascade-patterns of the top squark, the current lower bounds on its mass could also be relaxed. 
This is likely to offer an additional strengthening of the reach for the 
ewinos at the LHC beyond the nominal expectation because of a larger 
production rate of such a relatively light top squark (enjoying a relaxed lower bound on its mass). 
As a result, searches at the HL-LHC, in absence of a discovery, would extend 
the lower bounds on the respective masses in a maximal way. We explore this 
issue briefly at length in this work.

As for the Higgs-like scalars of the \z3nmssm, we are particularly interested 
in the relatively light singlet-like ones among them (in particular, $\as$) and the observed 
SM(doublet)-like Higgs boson while we consider the MSSM-like Higgs 
bosons ($A,H,\hpm$) to be much heavier and, hence, essentially decoupled. The latter 
is realized by considering a large value of $\alambda$. 
In fact, given that we already require a not so heavy singlino-like state which 
is the NLSP ($\ntrltwo \sim \msinglino$), the singlet-like (neutral) scalars, 
$\hs$ (with {\small $\mhssq \sim \msinglinosq$}; see
equation~\ref{eqn:cp-even-mass}) and
$\as$ (with {\small $\massq \sim \akappa \msinglino$}; see
equation~\ref{eqn:cp-odd-mass}) would only have masses in the same ballpark of
$\msinglino$.

However, note that, by restricting $\akappa$ to smaller values 
($\akappa < \msinglino$), one can find $\as$ to be pretty light and still 
passing existing experimental bounds when it is dominantly a singlet by nature,
which is the case since $m_A$ is taken to be large. Thus, $\as$ might turn out to be 
the lightest of the excitations that the scenario possesses.
Even if that be not the case, there are some general possibilities with $\as$ 
that are of particular phenomenological significance in the context of our 
present work. 
First, for $\mas \approx 2\mntrlone$, $\as$ would provide an efficient (and 
hence, maybe, much sought after) funnel for the mutual-annihilation of the 
moderately-tempered DM particle. Second, even when this is not the case, we 
would like to ensure that $\as$ is light enough such that the decay
$\ntrltwo \to \ntrlone \, \as$ is kinematically allowed. This would present us
with new possibilities of cascade decays of heavier neutralinos which are yet
to be considered by the LHC experiments in their searches and hence, when
considered, are likely to result in relaxed lower bounds on the masses of all the involved
SUSY states. As for
$\mhs \approx \msinglino$, one just needs to ensure that it is not too small 
such that the contribution of $\hs$ to the DMDD-SI rate does not result in a breach of 
its experimental upper bound on the latter.
\begin{table}[t]
\begin{center}
\begin{tabular}{|c|c|c|c|c|c|c|c|c|c|}
\hline
\makecell{Varying  \\ parameters}  & $\lambda$ & $|\kappa|$ & $\tanb$& \makecell{$|\mueff|$ \\ (TeV)}&  \makecell{$|\alambda|$ \\ (TeV)} &
\makecell{$|\akappa|$ \\ (GeV)}
 & \makecell{$|\mone|$ \\ (GeV)} & \makecell{$|A_t|$ \\ (TeV)} & \makecell{$m_{{\widetilde{Q}_3,\widetilde{U}_3}}$  \\ (TeV)} \\
\hline
 & & & & & & & & &  \\
Ranges  & 0.5--0.7 & $\leq$ 0.1 &
1--20 & $\leq 1$ & $\leq$ 10 & $\leq 150$ & $\leq$ 200 & $\leq$ 10 & 0.1--6 \\
 & & & & & & & & &  \\
\hline
\end{tabular}
\caption{Ranges of various input model  parameters considered in this work. The wino mass, $\mtwo$, is kept fixed at a value as high as 2.5~TeV.}
\label{tab:ranges}
\end{center}
\vspace{-0.5cm}
\end{table}

\begin{figure}[t]
\begin{center}
\includegraphics[width=0.58\linewidth]{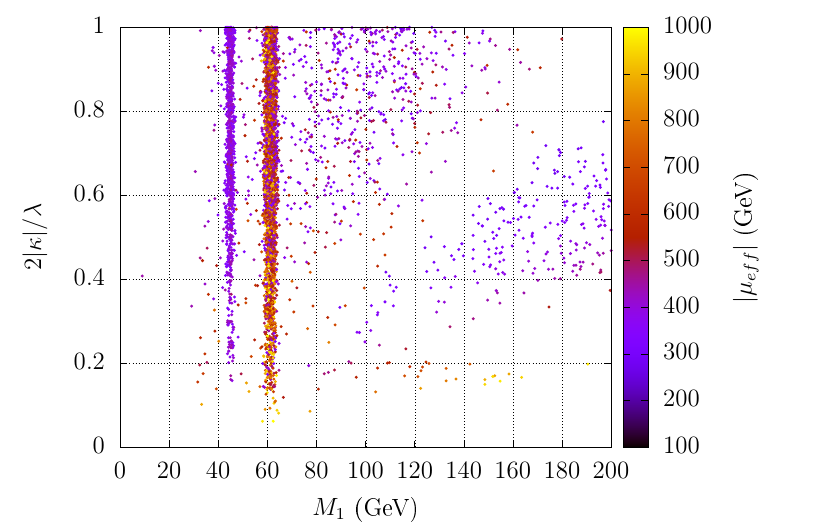}
\caption{Scattered points in the $M_1-2|\kappa|/\lambda$ plane that pass all relevant constraints from the DM and the Higgs sectors and those that are
implemented in \nmssmtools~(see section~\ref{subsec:constraints}). Colors in the adjacent palette are representative of the magnitude of $\mueff$ associated with each such point.}
\label{fig:scan_result_ewino}
\end{center}
\vspace{-0.5cm}
\end{figure}

Guided by the discussion above, in table~\ref{tab:ranges} we indicate the 
extended region of the \z3nmssm parameter space which would be interesting
for our purpose. 
A corresponding graphical illustration of such a motivated region of the parameter space is presented in figure~\ref{fig:scan_result_ewino}. There, the presentation  takes into account various experimental
constraints coming from the DM and the Higgs sectors and involves the three most crucial quantities, $M_1$, $2|\kappa|/\lambda$ and $\mueff$, that characterize the augmented ewino and the  Higgs sectors of the \z3nmssm. The criterion $2|\kappa|/\lambda < 1$ ensures the NLSP to be singlino-like when we choose the LSP to be bino-like, i.e., $|M_1| < 2|\kappa|/\lambda$.
The two vertical bands correspond to the region of the parameter space where a pair of bino-like DM particles annihilate via $Z$-boson and $\hsm$ funnels, respectively, whereas for all other points DM annihilation proceeds either through the $\as$-funnel or via its co-annihilation with the singlino-like NLSP.\footnote{The $\hs$ funnel is generally not an efficient mode of DM annihilation as it is $p$-wave suppressed.} 
Note that, bino-like LSP mass $\lesssim \mhsm/4$ is barely allowed since a light scalar with mass  $\lesssim \mhsm/2$, which might act as a funnel, attracts stringent constraints from the studies of decays of the $\hsm$ at the LHC. 
Furthermore, constraints coming from the DMDD sector (both SI and SD) and those from numerous searches of the ewinos at the LHC are continuously pushing $\mueff$ upwards. In such a situation, the LHC might not be sensitive to this region of parameter space since the production cross-section of a pair of a relatively heavier  higgsino becomes critically small. 
As discussed earlier, under such circumstances the larger production rate of a $\stone$ pair at the LHC could enhance the reach for such ewinos which would appear in the cascade decays of~$\stone$.

%
%
%
In section~\ref{sec:results}, we pick up a few representative (benchmark) 
scenarios out of the region of parameter space discussed above and study their implications for the LHC 
searches.
%
\section{The setup of the analysis}
\label{sec:setup}
In this section, we present the setup of our analysis. This includes a thorough description of various theoretical and experimental considerations ranging over the imposed constraints that are pertinent for the present work, the choice of the benchmark scenarios  which are in sync with and are representative of the motivated region of the \z3nmssm parameter space as outlined in section \ref{sec:param-space}, motivating the NMSSM specific signals in those scenarios, a discussion of the relevant SM backgrounds and a brief mention of the approaches we took to optimize the signal strength with respect to the background. In the process, we also outline the computational tools we employ to carry out a state-of-the-art analysis as warranted by a phenomenological study of the present kind.
\subsection{The spectra and the constraints}
\label{subsec:constraints}
We rely on the package
\nmssmtools{\tt -v5.5.3}~\cite{Ellwanger:2005dv,Das:2011dg} for generating the
\z3nmssm spectra. Viable spectra are required to be consistent with various 
known constraints, both of theoretical and experimental kinds, before those 
could be checked for their suitability for being called `benchmarks' for our 
purpose. While \nmssmtools~computes the masses, mixings and decays of various 
NMSSM excitations within the framework of \z3nmssm, the built-in package
\micromegas{\tt -v4.3}~\cite{Belanger:2006is}  provides us with the values of 
various key observables of the DM sector~\cite{Abdughani:2021oit}.

Furthermore, \nmssmtools~ensures, on the theoretical side, that the spectra are free of 
tachyonic states, do not give rise to Landau poles in the evolution of various 
couplings and that the resulting scalar potentials do not have unphysical 
global minima. On the experimental side, it checks for the constraints from 
the flavor sector and a few primary ones from the LEP, Tevatron and the LHC. 
Among the latter set of constraints, the lower bounds arising from the LHC 
experiments on the masses of the lighter ewinos and $\stone$ are 
particularly relevant for this work. In the context of the present study, the 
former restricts the masses of the higgsino-like ewinos (and hence 
$\mueff$) from below. To keep our basic proposition and hence the analysis 
simple, we consider all the sfermions (including the sleptons), except for 
$\stone$, to be rather heavy (in the multi-TeV regime). Thus, those excitations are 
effectively decoupled from the phenomenology we discuss in this work.
Thus, we make no attempt to reproduce the experimentally measured value of the anomalous muon magnetic moment ($\mu_{g-2}$)~\cite{Muong-2:2021ojo, Muong-2:2006rrc}
which generically requires relatively light smuons in a SUSY scenario.

On the other hand, constraints on the Higgs sector coming from recent LHC studies are checked with the help of the dedicated packages like
\higgsbounds{\tt -v5.8.0}~\cite{Bechtle:2020pkv} and
\higgssignals{\tt -v2.5.0}~\cite{Bechtle:2020uwn} which are interfaced with \nmssmtools. The mass of the (observed) SM-like Higgs boson is allowed to be in the range $122$ GeV $< \mhsm < 128$ GeV to factor in the estimated theoretical uncertainties in its prediction.  

As for the DM sector, in our conservative approach, we require the DM relic abundance ($\Omega h^2$) to fall strictly within the band 0.108 $< \Omega h^2 <$ 0.132 .
This amounts to allowing for a theoretical uncertainty of 10\% about the Planck-observed central value of 0.120~\cite{Planck:2018vyg}.
We also ensure compliance with various upper bounds on the rates of SI and SD DM scatterings off the target nucleus of dedicated experiments some of which are incorporated in \nmssmtools. Note that, these DM constraints rule out smaller $\mhs$, larger values of the coupling $g_{_{\hsm \ntrlone \ntrlone}}$ (mainly from DMDD-SI studies) and smaller values $\mueff$ (mainly for DMDD-SD studies)~~\cite{Badziak:2015exr, Cheung:2012qy, Cheung:2014lqa, Badziak:2015nrb, Badziak:2017uto, Abdallah:2020yag}.
%
\subsection{The benchmark scenarios}
\label{subsec:benchmarks}
In this section, we present four types of benchmark scenarios based broadly on 
the value of $\mstone$ while masses of the higgsino-like states ($\sim \mueff$) 
for each such category are so chosen that the intended decays of $\stone$ to 
the latter are kinematically allowed. Further, the phenomenology of top squarks 
is known to depend much on their `chiral' contents. Hence we present the 
limiting situations with $\stone \approx \stleft$ ($\thetastop \sim 0$) and 
$\stone \approx \stright$ ($\thetastop \sim \pi/2$), whenever warranted.
%

%
\afterpage{
\begin{table}[H]
\renewcommand{\arraystretch}{1.30}
\centering{
{\tiny\fontsize{7.8}{7.2}\selectfont{
\begin{tabular}{|c|@{\hspace{0.06cm}}c@{\hspace{0.06cm}}|@{\hspace{0.06cm}}c@{\hspace{0.06cm}}|@{\hspace{0.06cm}}c@{\hspace{0.06cm}}|c|@{\hspace{0.06cm}}c@{\hspace{0.06cm}}|}
\hline
\Tstrut
Input/Observables & \makecell{BP1} & \makecell{BP2} & \makecell{BP3} & \makecell{BP4} & \makecell{BP5}  \\
\hline
\Tstrut
$\lambda$   &  $0.588$ &  $0.588$ &  $0.689$               & $0.689$  & 0.662  \\
$\kappa$   &  $-0.052$  & $-0.052$ &  $0.047$   & $0.047$  & $0.096$ \\
$\tan\beta$        &  8.63   & 8.63   &  7.83      & 7.83 & 6.35\\
$A_\lambda$~(GeV)  &  6146    & 6146   &  $-7304$      & $-7304$ & 3112\\
$A_\kappa$~(GeV)   &  108.1  & 106.1  &  $-16.5$  & $-14.1$ & 121.4\\
$\mueff$~~(GeV)       &  645.5   & 645.5   &   $-936.6$   &  $-936.6$ & 452.4 \\
$M_1$~(GeV)        &  32.2    & 31.8    &  $62.5$    & 61.5 & -61.5 \\
$M_{\widetilde{Q}_3}$~(GeV)        &  600  & 5000 &  1100   & 6200 & 5300\\ 
$M_{\widetilde{t}_R}$~(GeV)        &  5500   & 670 &   5500  & 1200 & 100\\
$A_{t}$~(GeV)        &  6242  & 6242 &  4506    & 4506 & 4369\\ [0.05cm]
\hline
\Tstrut
$m_{\chi_1^0}$~(GeV)    &  31.9   & 31.6 &  61.9    & 60.9 & $-61.96$\\
$m_{\chi_2^0}$~(GeV)    &  $-108.5$   & $-108.5$ & $-131.6$    & $-131.9$ & 132.5\\
$m_{\chi_3^0}$~(GeV)    &  660.3  & 660.4   &  958.7   & 960.4 & 467.8\\
$m_{\chi_4^0}$~(GeV)    &  668.8   & $-668.9$  &  $-959.8$   & 961.6 & $-470.2$\\
$m_{\chi_5^0}$~(GeV)    &  2584.3   & 2586.7  &  2582.0  & 2587.4 & 2581.1\\
$m_{\chi_1^\pm}$~(GeV)  &  654.1  & 654.1 &  950.7  & 952.5 & 452.2\\
$m_{h_1} \equiv \mhsm$~(GeV)         &  122.8  & 124.1   &  126.1  & 124.8 & 125.6  \\
$m_{h_2}$~(GeV)         &  216.4    & 217.1   &  345.1   & 345.6 & 187.1\\
$m_{\as}$~(GeV)         &  64.5  & 63.8   &  65.1   & 66.6 & 65.1\\
$m_{H^{\pm}}$~(GeV)         &  5739.5   & 5732.6  & 7386.1   & 7383.8 & 2911.5\\ [1ex]
$m_{\tilde{t}_1}$~(GeV) & 1041.3 ($\approx \stleft$)    & 1027.4  ($\approx \stright$)   & 1438.1 ($\approx \stleft$)  & 1491.3 ($\approx \stright$) & 802.3 ($\approx \stright$) \\[0.12cm]
\hline
%
$\Delta m_1, \, \Delta m'_1$ (GeV) & $\sim 200, \, \sim 380$ & $\sim 190, \, \sim 370$ & $\sim 305, \, \sim 480$  & $\sim 355, \, \sim 530$ & $\sim 160, \, \sim 345$\\ [1.0ex]
$\Delta m_2, \, \Delta m'_2$ (GeV) & $\sim 430, \, \sim 465$ & $\sim 430, \, \sim 465$ & $\sim 700, \, \sim 745$  & $\sim 700, \, \sim 740$ &  $\sim 212 , \, \sim2 40  $\\
\hline
\Tstrut
$\Omega h^2$  &  0.115  & 0.113 &  0.117  & 0.121 & 0.017\\[0.10cm]
$\sigma^{\rm SI}_{\chi^0_1-p(n)} \times 10^{47}$~(cm$^2$)  &    0.16 (0.16) & 0.13 (0.14) &  1.56 (1.59)   & 1.62 (1.67) & 1.7 (1.8) \\[0.30cm]
$\sigma^{\rm SD}_{\chi^0_1-p(n)} \times 10^{43}$~(cm$^2$) & 7.4 (-5.8)  & 7.4 (-5.8)  & 1.38 (-1.1)  & 1.37 (-1.1) & 3.5 $(-2.7)$\\[0.25cm]

$\sigma^{\rm IS}_{b \bar{b}} \times 10^{31}$~(cm$^3/s$)  &    13.5 & 1.0  &  13.1   & 11.4 &  0.2 \\[0.30cm]
$\sigma^{\rm IS}_{c \bar{c}} \times 10^{31}$~(cm$^3/s$) & 0.03  & 0.05   & 0.3  & 0.5 & 0.01 \\[0.25cm]
$\sigma^{\rm IS}_{l \bar{l}, \{l = \mu^-,\tau^-\}} \times 10^{31}$~(cm$^3/s$) & 1.2  & 0.15   & 1.3  & 1.6& 0.02 \\[0.25cm]
\hline
	BR$(\tilde{t}_1\rightarrow \chi_1^0  t)$  & 0.04 & 0.2 & 0.03  & 0.2 & 0.15\\
 			BR$(\tilde{t}_1\rightarrow \chi_2^0 t)$  & 0.07 & 0.03 & 0.05   & 0.02 & 0.05\\
 			BR$(\tilde{t}_1\rightarrow \chi_3^0 t)$  & 0.45 & 0.18 & 0.45   & 0.19 & 0.17\\
 			BR$(\tilde{t}_1\rightarrow \chi_4^0 t)$  & 0.42 & 0.16 & 0.45   & 0.18 & 0.17\\
 			 			BR$(\tilde{t}_1\rightarrow \chi_1^\pm b)$  & 0.04 & 0.42 & 0.03  & 0.41 & 0.46\\[0.05cm]
        \hline
BR($\chi^\pm_1\to\chi_1^0 W^\pm $)  &  0.15 & 0.15 &  0.12   & 0.12 & 0.14\\[0.1cm]
BR($\chi^\pm_1\to\chi_2^0 W^\pm $)  &  0.85   & 0.86  &  0.88  & 0.88 & 0.86\\[0.05cm]
\hline
BR($\chi^0_2\to\chi_1^0 \as $)  &  1  &  1  & 1   & 1 & 1\\[0.05cm]
\hline
BR($\chi^0_3\to\chi_1^0 Z $)    &  0.1  & 0.1 &  0.07  &  0.08 &  0.05 \\[0.05cm]
BR($\chi^0_3\to\chi_1^0 \, \hsm $)  &  0.056 & 0.056 &  0.04   & 0.04 & 0.07\\[0.05cm]
BR($\chi^0_3\to\chi_2^0 Z $)    &  0.61   & 0.61 &  0.43  & 0.43 & 0.35\\[0.05cm]
BR($\chi^0_3\to\chi_2^0 \, \hsm$)  &  0.22  & 0.22 &  0.43   & 0.43 & 0.45\\[0.05cm]
\hline
\Tstrut
BR($\chi^0_4\to\chi_1^0 Z $)    &  0.06  & 0.06  &  0.04  & 0.04 & 0.1\\[0.05cm]
BR($\chi^0_4\to\chi_1^0 \, \hsm $)  &  0.09  & 0.09 &  0.08  & 0.08 & 0.03\\[0.05cm]
BR($\chi^0_4\to\chi_2^0 Z $)    &  0.25  & 0.25  &  0.44  & 0.44 & 0.52\\[0.05cm]
BR($\chi^0_4\to\chi_2^0 \, \hsm $)  &  0.57  &  0.57 &  0.43    & 0.43 & 0.24\\[0.05cm]
\hline
\Tstrut
BR($\as\to b \bar{b}$)  &  0.9 & 0.9 &  0.91  & 0.91 & 0.91\\[0.05cm]
\hline
\end{tabular}
}}
\caption{Benchmark scenarios allowed by all relevant theoretical and 
experimental constraints (see text for details). Shown are the values of the 
pertinent input parameters defining the scenarios and the resulting masses, 
mass-splits, mixings, branching fractions of the relevant states along with 
the values of various DM observables.}
\label{tab:benchmarks}
}
\end{table}
}

In table~\ref{tab:benchmarks}, we present five benchmark scenarios based on the above-mentioned criteria. The table 
contains, for each scenario, the input parameters that are varied, the 
relevant masses and mixings of various excitations, the branching fractions of 
the involved states decaying into some relevant ones along with the predicted 
values of some pertinent DM observables, such as the relic abundance and the
DMDD-SI and DMDD-SD scattering rates.

Scenarios BP1 and BP2 have $\mstone \gtrsim 1$ TeV while 
$\mueff \sim 650$ GeV. These values border on the reported lower bounds on the 
respective quantities from the LHC experiments, albeit somewhat naively. This is since,
as has been pointed out already, the decay patterns of the involved states in 
our benchmark scenarios differ from what have been assumed by 
the experiments while deriving the said bounds.\footnote{In that sense, we 
choose to remain conservative in the choice of these mass parameters and leave 
a quantitative estimation of the extent of possible relaxations of these bounds for a dedicated study.}
These two scenarios differ in the chiral contents of $\stone$: while in BP1,
$\stone \approx \stleft$, in BP2, $\stone \approx \stright$. On the other 
hand, scenarios BP3 and BP4 are chosen with somewhat heavier $\stone$ 
($\mstone \sim 1.4-1.5$ TeV) and higgsino-like ewinos
($\mueff \lesssim 950$ GeV) in mind. The lighter ewinos are also 
commensurately heavier in scenarios BP3 and BP4. These two points are chosen 
to demonstrate the sensitivity of the LHC to such more massive states in 
scenarios like ours. Here, the heavier such ewinos, which 
the LHC might find difficult to reach out to in their direct production modes, may still be produced in the decays of $\stone$.
Note that when we move from scenarios BP1 and BP2 to BP3 and BP4, only those modifications made in the input parameters that involve the top squark sector and $\mueff$ would have some
practical phenomenological implications. Also, again, in BP3 (BP4), $\stone \approx \stleft (\stright)$. 

It may be noted that the strongest possible lower bound on $\mstone$ ($\gtrsim 1.2$~TeV) as reported by some of the recent LHC analyses~\cite{ATLAS:2020aci} are based on a number of simplified assumptions (and that also, within the MSSM) which simply do not hold in specific but realistic scenarios like the one adopted in the present work wherein top squarks would undergo longer cascades  by virtue of novel such ones of ewinos. Hence any such bound is expected to get relaxed. A systematic analysis of such a relaxation is beyond the scope of the present work and we reserve the same for a future investigation. However, with such a possibility
in mind, we introduce one more scenario, BP5, with a comparatively small
$\mstone$ ($\approx 800$~GeV, with $\stone \simeq \stleft$) which might turn up lying just on the edge of exclusion in such a specific scenario under the same experimental analysis.

On the DM front, conformity 
with the Planck-observed relic abundance ($\Omega h^2$) is ensured via mutual 
annihilation of the LSP DM via $\as$ ($\hsm$) funnel in scenarios BP1, BP2 
(BP3, BP4 and BP5).
A compliance with the experimental upper bound on the DMDD-SI rate
($\sigma^{\mathrm{SI}}$) is achieved via the realization of a coupling `blind spot' ($g_{_{\hsm \ntrlone \ntrlone}} \sim 0$), which is 
aided by having $\kappa < 0 \, (> 0)$ when $\mntrlone$ and $\mueff$ carry same 
(different) signs as are the cases in scenarios BP1 and BP3 (BP2, BP4 and BP5).
These follow directly from the discussion of the blind spot  condition of 
equation~\ref{eqn:blindspot}.
As for the DMDD-SD rate ($\sigma^{\mathrm{SD}}$), it varies primarily 
as $1/\mueff^4$~\cite{Abdallah:2020yag}. This is palpable in the values presented for 
$\sigma_{\mathrm{SD}}$ across the benchmark scenarios in
table~\ref{tab:benchmarks}, all of which pass the latest experimental bound 
placed on it. 

\begin{figure}[t]
\begin{center}
\includegraphics[width=0.58\linewidth]{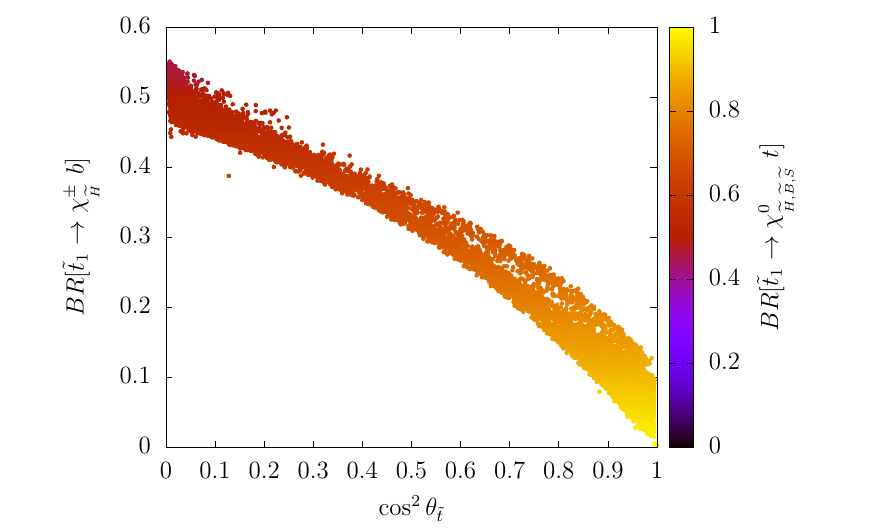}
\caption{Variation of $\text{BR}[\stone \to \chi^{\pm}_{\widetilde{H}} \, b]$ as a function of the mixing parameter of the stop sector, i.e., $\cos^2\theta_{\tilde{t}}$. The corresponding combined decay branching fraction of $\stone$ to the lighter ewinos is shown via the palette.}
\label{fig:scan_result_stop}
\end{center}
\vspace{-0.5cm}
\end{figure}
%

Turning to the relevant decays that shape the phenomenology at the LHC, one 
finds that, in all the benchmark scenarios, $\stone$ decays dominantly to the 
higgsino-like states (albeit being heavier of the accessible ewinos) 
thanks to the large $y_t$-driven interactions. Thus, as discussed in
section~\ref{subsubsec:ewino-stop1}, for
$\stone \approx \stleft \, (\stright)$, the decays
$\stone \to t \, \ntrlthreefour \; (\stone \to b \, \charonepm)$ 
dominate, which is the case with scenarios BP1 and BP3 (BP2, BP4 and BP5).
Such an observation is further in line with the general pattern of variations of the relevant decay branching fractions of $\stone$ with the mixing parameter of the stop sector, i.e., $\cos^2\theta_{\tilde{t}}$, as resulted from our scan, which is illustrated in figure~\ref{fig:scan_result_stop}.
Such a  well-defined, falling (with $\cos^2\theta_{\tilde{t}}$) band has its origin in  low to moderate $\tan\beta \lesssim 20$  that we consider in our scan. In fact, for larger values of $\tan\beta$ when $y_b$ gets enhanced, we would expect large values of $\text{BR}[\stone \to \chi^{\pm}_{(\widetilde{H})} \, b]$ (up to $\sim 0.5$) even for large $\cos^2\theta_{\tilde{t}}$, when $\stone \to \stleft$.

The size of the decay branching fractions of the ewinos to their lighter cousins, 
accompanied by a Higgs (scalar) or a gauge boson, following directly from the 
discussions in section~\ref{subsubsec:ewino-higgs-gauge}. As noted there, the 
decays of the higgsino-like $\ntrlthreefour, \charonepm$ to a singlino-dominated
$\ntrltwo{_{(\singlino)}}$ or to a bino-dominated $\ntrlone{_{(\bino)}}$ depend on the higgsino 
admixtures in the latter states and, for a larger `$\lambda$', these are larger for $\ntrltwo{_{(\singlino)}}$. This is the reason why, in 
table \ref{tab:benchmarks}, we find $\ntrlthreefour$ dominantly decaying to 
$\ntrltwo$ along with a $Z$-boson or a $\hsm$. Further, in scenarios BP1 
and BP2, the observation that the decay of $\ntrlthree$ dominantly yields 
$Z$-boson while that of $\ntrlfour$ has a preference for $\hsm$ is attributed 
to the so-called antipodal behavior of $\ntrlthree$ and $\ntrlfour$~\cite{Abdallah:2020yag, Calibbi:2014lga} in this regard.
This effect is less dominant in scenarios BP3, BP4, and BP6 when 
$\ntrlthreefour$
are much purer higgsino states given that larger values 
of $\mueff$ in these scenarios result in smaller mixings among the higgsinos and the singlino.
The scenario BP5 also shows cases of such an antipodal pattern in the decays of
$\ntrlthreefour$, albeit the preferred modes of their decays get flipped when 
compared to scenarios BP1 and BP2. Also, note that the mass-splits between the 
singlino-dominated $\ntrltwo$ and the bino-dominated $\ntrlone$ are such that 
BR$[\ntrltwo \to \ntrlone \, \as]=1$ for all the scenarios presented in 
table~\ref{tab:benchmarks} and in all these cases $\as$ mostly decays to a 
$\bbbar$ pair.

A pertinent observation at this point is that both $\hsm$, originating in the 
decays $\ntrlthreefour \to \ntrltwo \, \hsm$, and $\as$, arising in the decay
$\ntrltwo \to \ntrlone \, \as$, could be reasonably boosted.
This can be understood in terms of various mass-splits among the states taking part in the cascade decays of $\stone$ of which the following four, down the cascade, are relevant in our setup:
(i) $\Delta m_1 = \mstone - (\mntrlthreefour + m_t)$,
(ii) $\Delta m'_1 = \mstone - (\mcharone +m_b)$ and
(iii) $\Delta m_2 = \mntrlthreefour - (\mntrltwo + \mhsm)$,
(iv) $\Delta m'_2 = \mcharone - (\mntrltwo + m_\wpm)$.
These sets of values are indicated for each of the benchmark scenarios in table~\ref{tab:benchmarks}.
While $\ntrltwo$ and $\hsm$, having similar masses, would share the overall boost 
available to them by virtue of the magnitude of $\Delta m_2$, the absolute boost 
acquired by $\hsm$ would always be significant for a larger $\Delta m_2$, i.e., 
for larger $\mntrlthreefour \sim \mueff$. Nonetheless, the boost originally 
inherited by $\ntrlthreefour$ from the decays $\stone \to \ntrlthreefour t$ at the 
top of the cascade and which depends upon $\Delta m_1$, does percolate to some 
extent to $\hsm$ downstream. Hence, for a fixed $\mstone$ and $\mntrltwo$, as 
$\mntrlthreefour$ decreases (i.e., $\Delta m_1$ increases), the boost inherited by $\hsm$ from the top of the 
cascade would likely increase at the cost of the boost it acquires immediately in the 
decays $\ntrlthreefour \to \ntrltwo \, \hsm$ (as $\Delta m_2$ deceases).

On the other hand, $\as$ is likely to derive most of its boost from that of 
$\ntrltwo$ (and hence, again, the boost would mostly depend on $\Delta m_2$)
as it decays, thanks to the states participating in the decay 
$\ntrltwo \to \ntrlone \as$ are all packed closely in mass at the bottom of the 
spectra.
However, note that $\ntrltwo$, and hence $\as$, arising from the cascade
$\stone \to t \ntrlthreefour, \; \ntrlthreefour \to \ntrltwo Z$ or from the 
cascade $\stone \to b \charonepm, \; \charonepm \to \ntrltwo W^\pm$ would be 
further boosted.
This is primarily due to the fact that $m_{Z/W^\pm} < \mhsm$, which has two implications
for an enhanced boost for $\ntrltwo$. First, both $\Delta m_2$ and $\Delta m_2'$ are now larger thus resulting in larger overall boosts that would be available to the $Z/W^\pm \ntrltwo$ final states (arising in the decays of $\ntrlthreefour/\charonepm$) and (ii) $\ntrltwo$ would now carry a bigger share of the available boost as compared to when it arises in association with $\hsm$ in the decays of $\ntrlthreefour$.

In summary, in our benchmark scenarios, $\hsm$ and $\as$ would be reasonably 
boosted because of much heavier $\ntrlthreefour \sim \mueff$ while
$\mntrlonetwo \sim \mone,\msinglino$ being on the lighter side. Hence the 
angular separations between the two $b$-jets originating from decays of  $\hsm/\as$ 
would be on the smaller side, thus possibly giving rise to fat jets 
(which we call the `Higgs jets') which could help identify such 
scalar states.
%
%
\subsection{An NMSSM-specific signal and the SM backgrounds}
\label{subsec:signal-bg}
A suitable NMSSM-specific signal under cascade decays of a pair-produced $\stone$ in 
our target scenario could be figured out right from the decay patterns of $\stone$ and the ewinos as illustrated in the Introduction. As indicated there, $b$-jets arising 
from the decays of the top quarks and various Higgs bosons ($\hsm$ and $\as$) 
produced from the cascading $\stone$ or from its direct decay to
$b \, \charonepm$, which is predominant for $\stone \sim \stright$, would 
dominate the signal final states. Note that the latter mode of decay of 
$\stone$, in which $\charonepm$ appears in place of  $\ntrlthreefour$, would 
subsequently miss out on an $\hsm$ down the cascade in favour of a $W^\pm$
boson.  

In a rather busy final state that is thus expected under the circumstances, we 
lean heavily on an early tagging of the light NMSSM state $\as$ ($\mas < m_Z$) 
by reconstructing the same in the peak of the invariant mass of a $b$-jet pair  which it would decay dominantly to. In addition, tagging at least one $\hsm$ at the invariant mass-peak of another $b$-jet pair would prove helpful in 
dealing with the background. Requiring an isolated lepton ($e,\mu$) with its 
origin in the decay of a $W^\pm$ (coming from the decay of a top quark or a
$\charonepm$ to which a $\stone$ decays to at the top of the cascade) would 
help tame the mighty QCD background. As usual, the escaping LSPs would be the 
dominant contributors to $\etmiss$ and an optimally strong lower cut on it is 
indispensable to suppress the SM background. Furthermore, in each event, there 
could be up to 6 additional $b$-jets that arise from decays of the top quarks 
and the untagged $\hsm$ and $\as$ in the event. In addition, there will be a 
pair of light quark jets coming from the other $W^\pm$ boson present in the 
event. Consequently, the final states will still be rather rich in jets (up to 
8, at the parton level) even after the reconstruction of one each of $\as$ and $\hsm$. Thus,
the signal we choose to study in this work is as follows: 
\beq 
 1\ell \, (e,\mu) + (\geq 1)~ a_s \, (\bbbar) + (\geq 1)~ \hsm \, (\bbbar)+
\geq \, 4 \; \mathrm{jets}  + \etmiss \; .
\label{eqn:signal}
\eeq
Note that requiring at least one $\hsm$ in the signal makes the latter draw 
contributions only from the cases where either one or both $\stone$'s decay to
$t \ntrlthreefour$. In the first case, the other $\stone$ would need to decay 
to $b \charonepm$ for the purpose.

The dominant SM backgrounds for this  final state
 arise from processes involving a $\ttbar$ pair, viz.,
\beq
p p\to \ttbar, \, \ttbar \hsm, \, \ttbar \bbbar, \, \ttbar Z, \, \ttbar W^\pm \, ,
\label{eqn:smbg}
\eeq
where, the invariant mass of various pair-wise combinations of $b$-jets arising in the decays of the top quarks, $Z$-boson and $\hsm$ might get reconstructed to $\mhsm$ and/or $\mas$ thus faking the signal.
On the other hand,
possible backgrounds from processes like $pp \to W^{\pm}+\text{jets}, Z+\text{jets}, \hsm+\text{jets}, W^\pm \hsm, Z\hsm, W^\pm Z$ are expected to be negligible~\cite{ATLAS:2022ihe, ATLAS:2021twp} when the signal contains so many hard jets alongside a hard lepton.
%
\subsection{The simulation framework}
\label{subsec:framework-analysis}
Events for both the signal and the backgrounds are generated by considering up to 
two extra parton-level jets for each of the primary processes the primary processes contributing at the
lowest order (LO) in the perturbation theory using
{\tt MadGraph5-aMC$@$NLO-v2.7.3}~\cite{Alwall:2014hca} for the 13 TeV 
LHC run. The default parton distribution function
({\tt `nn23lo1'}~\cite{Ball:2010de}) is employed 
 with a dynamic choice of the factorization scale
given by $Q^2= \frac{1}{2} \sum_i M_{i}^{T^2} =\frac{1}{2}\sum_i(m_i^2+p_{Ti}^2)$, where $M^T_i$ is the transverse mass of the `$i$'-th final state particle.
The primary process contributing to the signal is $pp \to \stone \stone^*$, while
for the backgrounds, those are the ones mentioned in equation~\ref{eqn:smbg}.
To generate the signal events {\tt MadGraph5} is interfaced with the NMSSM 
{\tt UFO}~\cite{Degrande:2011ua} model file as obtained from the package
{\tt FeynRules}~\cite{Alloul:2013bka}. 

Computations of the decay-kinematics of the unstable excitations and 
subsequent showering and hadronization are carried out using
{\tt PYTHIA8-v8.3}~\cite{Sjostrand:2006za,Sjostrand:2007gs} using the default
\pythia8 cards.
As for the model-dependent masses and decay branching fractions of various NMSSM excitations, \pythia8 reads those in from the {\nmssmtools-generated} SLHA \cite{Skands:2003cj} files linked to it for the purpose.
We employ the MLM~\cite{Mangano:2006rw} approach, as implemented in {\tt MadGraph5}, for the purpose of matching of partonic jets (from the LO matrix elements) to those coming from the parton showers. Finally, detector responses are taken into account using
{\tt DELPHES-{v3.4.2}}~\cite{deFavereau:2013fsa} running with the 
CMS detector card with some modifications in its lepton isolation criteria which are discussed in section \ref{subsec:objreco}.

Radiatively corrected cross section for the base process, i.e.,
$pp \to \stone \stone^*$,  contributing to the signal is considered at the 
NLO+NLL level by employing a flat $k$-factor of 1.4~\cite{Broggio:2013uba}.
As for the base background processes, the $k$-factors 
considered are 1.4~\cite{Melnikov:2009dn}, 1.2~\cite{Beenakker:2002nc}, 1.8~\cite{Buccioni:2019plc},
1.35~\cite{Kardos:2011na}, 1.5~\cite{vonBuddenbrock:2020ter} for the 
processes
$pp \to \ttbar, \, \ttbar \hsm, \, \ttbar \bbbar, \, \ttbar Z, \, \ttbar W^\pm$, respectively.

\subsection{Object reconstruction}
\label{subsec:objreco}
In this section, we describe how various composite observables are constructed out of the primary objects mentioned above, once the detector effects are considered via {\tt DELPHES}.
\begin{itemize}
\item
{\bf Missing Transverse Energy (MET or $\etmiss$):} The missing $\pt$ vector is 
constructed out of the negative of the resultant $p_T$ of all the visible 
entities in the detector and its magnitude is used as $\etmiss$.
\item
{\bf Lepton ($e$, $\mu$) selection:} Non-isolated electrons and muons are 
selected using the following $\ptell$ and (pseudo)rapidity ($\etaell$) cuts:
\beq
\ptell > 30~\mathrm{GeV}, \quad |\etaell|<2.5 \;, \;\; \ell=e,\mu \;,
\eeq
while mini-isolation~\cite{CMS:2016krz} of each candidate lepton is ensured by demanding the 
combined $\pt$ of all objects within a cone of radius $r=10/\pt^{(\ell)}$ about the candidate 
lepton is less than a certain fraction of the $\pt$ carried by the same and 
hence imposing the condition
\beq
\frac{\sum \pt^{(\deltar<r)}}{\ptell}<I(\ell) , 
\eeq
where $\deltar=\sqrt{\Delta_\phi^2+\Delta_\eta^2}$, $\Delta_\phi$ and 
$\Delta_\eta$ being the angular distances from the candidate lepton in the 
azimuthal and the beam directions, respectively, and $I(e,\mu)=(0.12,0.25)$. 
\item
{\bf Reconstructing the SM-like and the light (singlet-like) Higgs bosons:}
As discussed earlier, due to their likely boosted nature, 
these Higgs bosons are reconstructed as fat (Higgs) jets. The low-level energy flow (e-flow) objects, 
e.g., the tracks, the photons and the neutral hadrons are used as inputs to the 
package {\tt Fastjet}~\cite{Cacciari:2011ma}. Jets are then constructed using the
Cambridge-Aachen (CA) algorithm with the radius parameter set to $R=1$ and by 
requiring $\ptjet>100$ GeV. The fat jets thus obtained are then 
treated with the mass drop tagger (MDT)~\cite{Butterworth:2008iy,Dasgupta:2013ihk}, with $\mu=0.67$ and 
$y_{_{\mathrm{cut}}}>0.09$, 
to get rid of the $b$-jets from gluon splittings and the contamination from 
the underlying events (UE) while retaining the $b$-jets from the decays of the Higgs bosons and 
accompanying hard radiations off them~\cite{Butterworth:2008iy}. The 
fat jets that pass MDT are referred to as tagged fat jets ($\text{FJ}_{b\bar{b}}$).

\parindent 20pt
The two subjets unearthed during the first stage of declustering of each 
tagged fat jet are then matched with the $b$-quarks of a given event within a 
matching cone $\deltar = 0.3$. When both subjets are found to be 
$b$-like, the covering fat jet is identified as either $\hsm$ or $a_{s}$ 
depending on whether $m_{_{\mathrm{FJ} {(bb)}}}\in[100,150]$ GeV or
$m_{_{\mathrm{FJ} {(bb)}}}\in[50,80]$ GeV, the latter window being designed 
with the lighter Higgs boson in mind that has a mass of around 65 GeV for the benchmark 
scenarios we consider in this work.
\item 
{\bf Reconstructing the ordinary jets:} After reconstructing $\hsm$ 
and $\as$, the e-flow objects that are away from all the jets 
that constitute those Higgs bosons are clustered to form the ordinary jets using 
{\tt Fastjet} and by employing the anti-$k_t$ algorithm, with the jet radius 
parameter $R=0.5$ and $\ptjet >20$ GeV. To identify such jets with possible 
residual $b$-jets in the event (for the signal, the ones from the top quark 
decays are there for sure), these are again matched with the $b$-quarks of the event.
\end{itemize}
\section{Results}
\label{sec:results}
With the setup of our analysis now ben described, we move on to present our results. As mentioned earlier, two standard 
approaches are adopted to optimize the signal to background ratio, i.e., (i) the 
usual cut-based analysis (CBA) and (ii) the MVA using the BDT 
which learns and exploits the intrinsic kinematic features of the signal 
and the backgrounds (that may be essentially inaccessible to the CBA) thus aiding a more efficient discrimination of the signal from the 
background.  It may be noted that LHC  analyses triggering on a single lepton would include the kind of final state (equation \ref{eqn:signal}) that has been considered in this work.
\subsection{The cut-based analysis}
\label{subsec:cut-based}
The cut-based analysis, for our purpose, banks primarily on the following kinematic variables:
\begin{itemize}
\item $m_{_{\mathrm{FJ} {(bb)}}}$, the invariant mass distribution  of a pair of $b$-jets constituting a fat jet thanks to their large boosts when those originate in the decays of $\hsm$ and/or $\as$, as is the case for the signal final state,
\item $m_{_{T}} (p^{({\ell})}_T, \etmiss)$, the transverse mass of a lepton-$\etmiss$ system, conventionally given by
\beq
m_{_{T}} (p^{({\ell})}_T, \etmiss) = \sqrt{2\times p_T^{(\ell)}\times \etmiss \times (1-\cos\Delta\phi)} \; , \nonumber
\label{eq:mtbb}
\eeq
where $\Delta \phi$ is the opening angle between a charged lepton and the 
direction of $\etmiss$ in the azimuthal plane. Given that $\etmiss$ in a signal 
event is not directly correlated with the kinematics of the final-state lepton 
unlike in the case of background events where the leptons and the neutrinos, the sole carriers 
of $\etmiss$ there at the parton level,  both originate in the 
decays $W^\pm \to \ell^\pm \nu_{_{\ell}}$, this variable is expected to serve as
a good discriminator between the signal and the background events.
\item The variable $H_T^{(j)}$, defined as the scalar sum of transverse momenta 
of all the jets (including the $b$-jets), i.e.,
$H_{T}^{(j)} = \mathlarger{\sum}_j \left| p_T^{(j)} \right|$ ,
is expected to peak at and extend to much higher values for the signal when compared to 
the background. This is since, for the former, the jets originate in the 
cascades of rather heavy SUSY states and, at the same time, the
jet-multiplicity is much higher in our chosen signal final state than that is 
expected from the background. Note that, in our (cut-flow) study, only those 
jets that are left out after the reconstruction of $\hsm$ and $\as$ concern 
$H_T^{(j)}$.
\item  The variable $R_n$, which is the ratio of the summed $p_T$ of the leading  `$n$' `ordinary' jets and that involves all `ordinary' jets in a given event and is given by \cite{Guchait:2011fb}
\beq
R_n = \frac{\sum_{j=1}^{n} \left| p_{T}^{(j)} \right|}
{
H_T^{(j)}} \, \,. \nonumber
\eeq
Operationally, in this work, the `ordinary' jets are comprised of all possible jets, 
including the $b$-jets that are left in an event after the reconstructions of 
$\hsm$ and $\as$. By construct, $R_n=1 (< 1)$ for the events with
$N_{\mathrm{jet}} \leq n (>n)$. Our choice of $n=3$, i.e., of $R_3$, and a 
nontrivial lower cut on it, are guided by the fact that the signal final states 
in our scenario are expected to be characteristically rich in more than three rather hard jets, unlike in the case of the background. 
\item $\etmiss$; last but not the least, this variable, as is broadly the case
in the search for physics beyond the SM with a new neutral, heavy, and stable 
excitation, remains to play the role of a crucial discriminator between a new 
physics signal and the SM background having typically large values for the former.
\end{itemize}
We now take a closer look at the differential distributions of both the 
signal and the background events in all these kinematic variables to decide 
over an optimal set of cuts on them that tends to maximize the signal 
significance over the background.

\begin{figure}[t]
\begin{center}
\subfigure[]{
\includegraphics[width=0.46\textwidth,height=0.3\textheight]{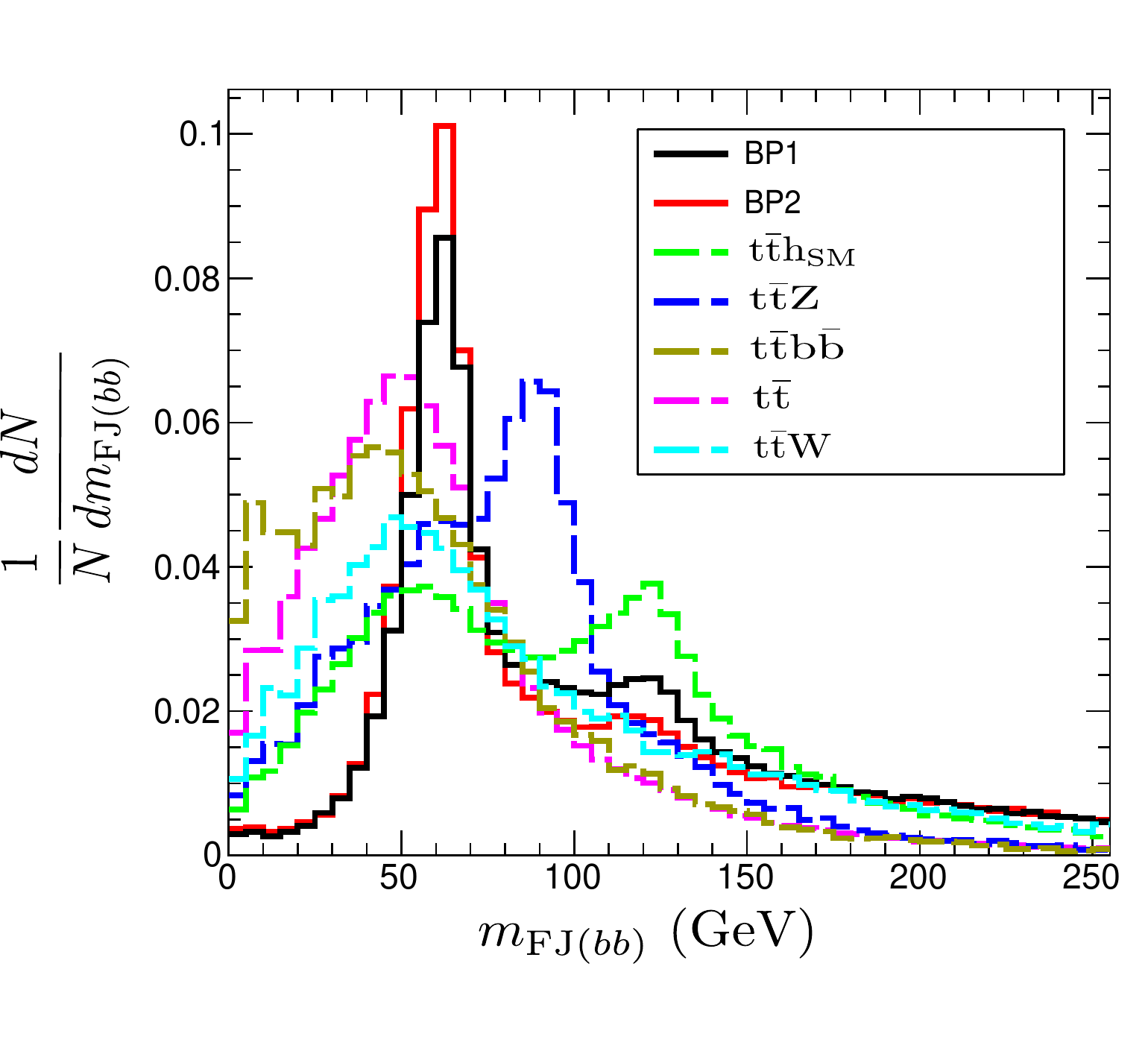}
\label{subfig:mbbbar}
} \\
\subfigure[]{
\includegraphics[width=0.46\textwidth,height=0.3\textheight]{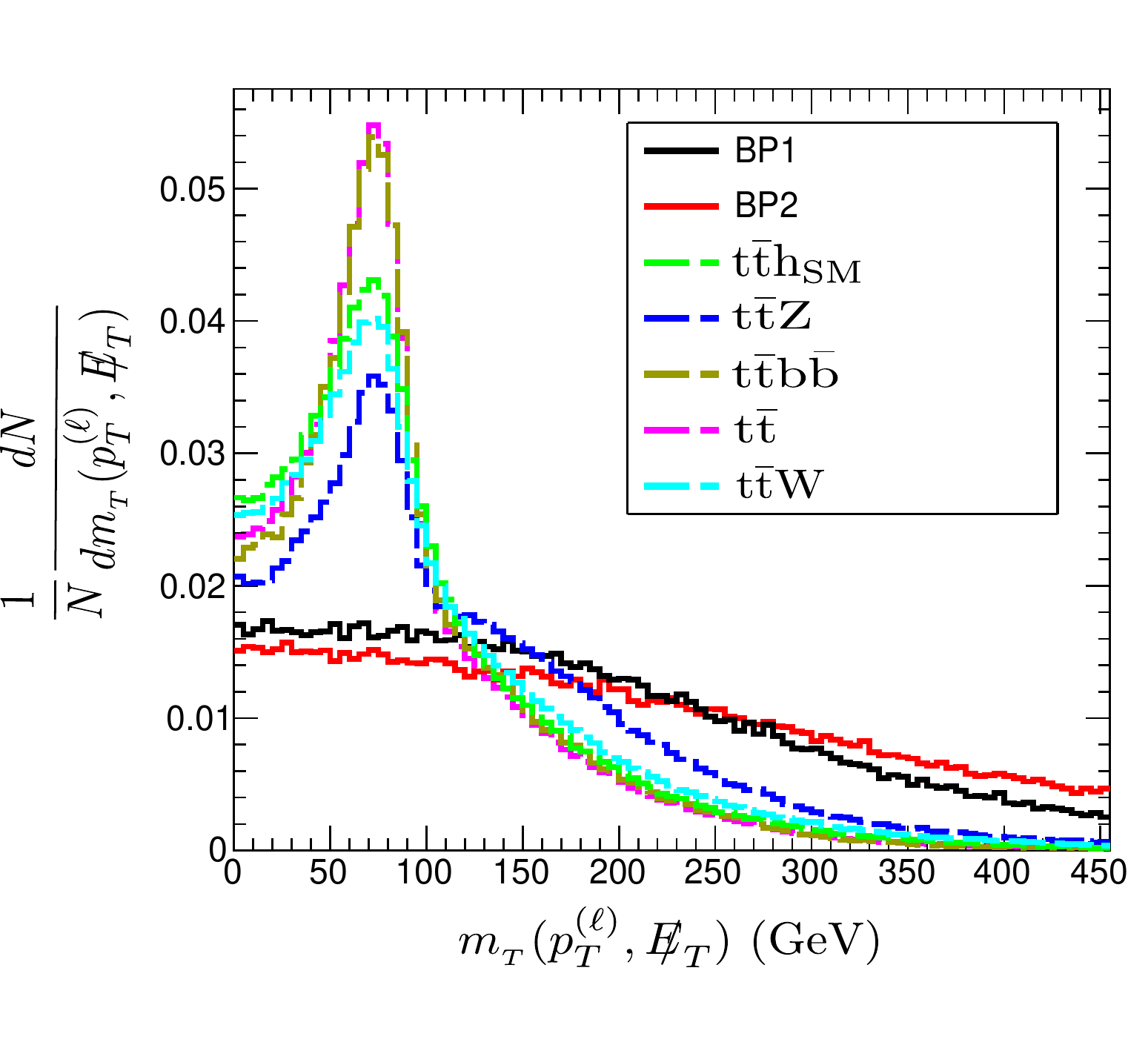}
\label{subfig:mtransverse}
}
\hskip 15pt
\subfigure[]{
\includegraphics[width=0.46\textwidth,height=0.3\textheight]{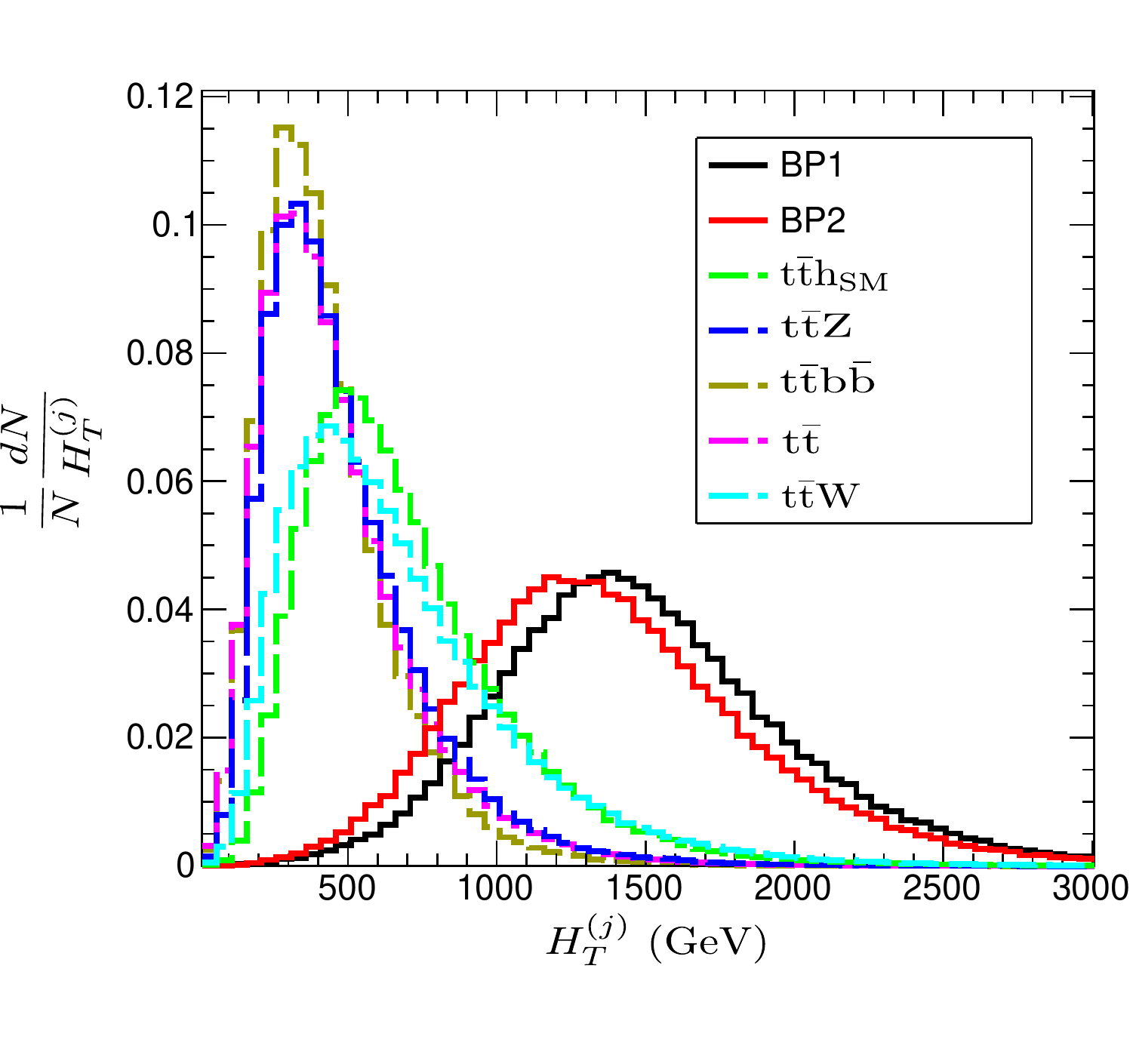}
\label{subfig:mhtjet}
}
\end{center}
\caption{Normalised differential distributions of the number of events in 
the kinematic variables
(a) $m_{_{\mathrm{FJ} {(bb)}}}$, (b)~$m_{_{T}} (p^{({\ell})}_T, \etmiss)$ 
and (c) $H_T^{(j)}$ for the backgrounds from various SM processes and for the 
signals in the  benchmark scenarios BP1 and BP2.
}
\end{figure}

In figure~\ref{subfig:mbbbar} we present the invariant mass
($m_{_{\mathrm{FJ} {(bb)}}}$) distributions of pairs of $b$-jets where the 
$b$-jets in each such pair are required to be found as a fat jet, for both the 
signal and the prominent backgrounds. Scenarios BP1 and BP2 are chosen as the 
representative ones for presenting the signal distributions. Three peaks are 
discernible in these distributions. The ones seen at around 65 GeV solely 
appear for the signal and indicate the presence of $\as \, (\to \bbbar)$ in 
the event. The peak in the middle around $m_Z$ comes from the background 
process $pp \to \ttbar Z$ with the $Z$-boson decaying to a $\bbbar$ pair. The 
peaks on the right occur around $\mhsm$ and have their origins in the decays
$\hsm \to \bbbar$ in the signal where $\hsm$ appears in the cascade decays of 
$\stone$ and in the background from the process $pp \to \ttbar \hsm$. Relatively
taller peaks at about $\mas (\approx 65$ GeV) when compared to those about 
$\mhsm$ solely reflect larger  yields of $\as$ than that
of $\hsm$ in the signal thanks to larger effective branching fractions  in the former mode. It is 
found that, for the scenarios of our choice, requiring the events to lie
in the window $|m_{_{\mathrm{FJ} {(bb)}}}-\mas| < 15$ GeV picks up the 
signal events containing an $\as$ rather efficiently. Furthermore, selecting 
simultaneously the window $|m_{_{\mathrm{FJ} {(bb)}}} -\mhsm| < 25$ GeV 
helps significantly in getting rid of the background coming from the SM process
$pp\to  \ttbar \hsm$. Qualitatively, at the level of the
$m_{_{\mathrm{FJ} {(bb)}}}$ distributions, the overall situation does not change 
much in going from scenarios BP1 (BP2) to scenarios BP3 (BP4), respectively.

In figure~\ref{subfig:mtransverse} we present the differential 
distributions of the events in the variable $m_{_{T}} (p^{({\ell})}_T, \etmiss)$ for the same 
set of background processes as done for figure~\ref{subfig:mbbbar}. It can be seen 
that for all the 1-lepton final states contributing to the background which 
arise from processes that include a $\ttbar$ pair, the distributions of 
$m_{_{T}} (p^{({\ell})}_T, \etmiss)$ fall sharply beyond the (Jacobian) peaks 
about $m_W$. A visible departure (flattening) from such behavior over the 
tail, however, is observed for the process $pp \to \ttbar Z$ which is to be 
attributed to the additional contributions to $\etmiss$ coming from the decays
$Z \to \nu\bar{\nu}$ that are uncorrelated to the $\ell \nu$-system arising from
$W^\pm$-boson decays. In general, for all the background processes, the extended
tails in the $m_{_{T}} (p^{({\ell})}_T, \etmiss)$ distributions (instead of a 
more sharply dropping behavior, signifying an endpoint at $m_W$, as is 
theoretically expected) are to be attributed to the detector-induced smearing 
of momenta of various final state objects, which affects the estimation of 
$\etmiss$ and hence $m_{_{T}} (p^{({\ell})}_T, \etmiss)$. The accumulated 
effect becomes somewhat pronounced in the presence of more number of jets in 
the final state, as is the case with our signal.
On the other hand, a similar lack of a strong correlation 
between the only lepton and $\etmiss$ in the signal final state makes the
corresponding $m_{_{T}} (p^{({\ell})}_T, \etmiss)$ distributions much too flatter. Further, 
those distributions extend to rather high values of $m_{_{T}} (p^{({\ell})}_T, \etmiss)$ as both the lepton
and $\etmiss$ in a signal event could get to be much harder. Guided by these
distributions, we require $m_{_{T}} (p^{({\ell})}_T, \etmiss) \geq 100$ GeV to 
further optimize the signal to background ratio, once at least one each of $\hsm$ and $\as$ are tagged.

Figure~\ref{subfig:mhtjet} 
illustrates the corresponding distributions for the variable $H_T^{(j)}$. As 
pointed out earlier, the peaks of the signal distributions shift to higher values
when $\mstone$ increases and for the two benchmark scenarios these
peaks are already well-separated from those for the background processes occurring 
at lower values of $H_T^{(j)}$. Given these, choosing
$H_T^{(j)} \geq 300$ GeV is found to be optimal for enhancing the signal 
to background ratio,  once at least one each of $\hsm$ and $\as$ are tagged.

As for the variable $R_3$, we impose a cut of $R_3 < 0.85$ to maximize 
the signal significance. This implies that a $p_T$-sum amounting to more than 
15\% of $H_T^{(j)}$ has to be carried by the jets in the events which are
present beyond the three leading ones. As for a cut on $\etmiss$, we impose a 
moderately hard cut of $\etmiss > 100$ GeV at the very outset. However, 
given that the variable $m_{_{T}} (p^{({\ell})}_T, \etmiss)$ is closely correlated to 
$\etmiss$, a subsequent lower cut on the former makes any further lower cut on the latter an overkill and hence unnecessary.

In table~\ref{tab:cross-section} we present the cut-flow of the yields for our benchmark scenarios and for the primary SM background 
processes. The LO cross-sections of each of the
processes, computed using {\tt MadGraph} (with its default kinematic cuts), for the centre of mass energy of $\rm \sqrt{s}=13~TeV$ LHC are presented in the first column. The imposed cuts are what have been discussed earlier and are 
indicated in the top row of the table. These lead to the effective yields ($\sigma_{eff}$), presented in the last 
column of the table. These are obtained by multiplying the fiducial cross-section $\sigma_{\mathrm{fid}}$ by the appropriate
$k$-factors (as discussed in section~\ref{subsec:framework-analysis}) to take into account the higher-order corrections to the respective cross sections.
It may be reiterated that requiring a significant $\etmiss$ ($ > 100$ GeV) and a 
lepton in the events, at the very outset, would drastically suppress the large 
QCD background. However, the significant direct backgrounds from the processes 
involving a $\ttbar$ pair, as can be gleaned from the entries in table~\ref{tab:cross-section}, could be tamed rather effectively in our studies initially by requiring $\hsm$ and $\as$ reconstructed and subsequently by putting cuts on $m_{_{T}} (p^{({\ell})}_T, \etmiss)$, $H_T^{(j)}$ and $R_3$. It is seen that while the cuts on $m_{_{T}} (p^{({\ell})}_T, \etmiss)$, $H_T^{(j)}$ affect the backgrounds moderately, the cut on $R_3$ drastically
eliminates them almost without hurting the signal.

Table~~\ref{tab:signalsignificances} summarizes the signal significances
$\sigma = S/\sqrt{S+B} \,$, where `$S$' and `$B$' correspond to the number of
signal and the background events, respectively, after all the cuts are imposed,
for all the benchmark scenarios and for integrated luminosity values of 
300~\fbinv~and 3000~\fbinv~at the LHC. As can be seen, with
${\cal L}=300$~\fbinv~at the LHC, even by exploiting the cascades of $\stone$, the lighter
of the ewinos which the LHC would not be able to reach out to in their direct searches, cannot be even remotely probed. An integrated luminosity of 3000~\fbinv, for $\mstone \gtrsim 1$ TeV, as are the cases with scenarios BP1 and BP2, could scale
(as $\approx \sqrt{\cal L}$) the
signal significance up above the coveted $5\sigma$ level when one could expect to probe those lighter ewinos along with an NMSSM-specific lighter scalar in the cascade decays of $\stone$. However, for all other scenarios (like BP3 and BP4),
with $\stone$ moderately heavier than what they are in cases BP1 and BP2, the signal significance still remains abysmally small to be of any practical use, even for ${\cal L}=3000$~\fbinv. We further note that 
these signal sensitivities can get reduced by $\sim$ 0.01\% if one considers a flat 10\% uncertainty in the estimation of the SM background.
%

%
\begin{table}[t]
\renewcommand{\baselinestretch}{1.4}
\begin{center}
{\tiny\fontsize{8.0}{8.0}\selectfont{
\begin{tabular}{|c|c|c|c|c|c|c|c|c||c|}
 \hline
   {\makecell{Signal \\ scenario}} 
& {\makecell{$\sigma$ \\ (LO)}} & {\makecell{$\etmiss$ \\ $ > 100$ GeV}}
 & {\makecell{$n_\ell$ \\ $=1$}} 
 & {\makecell{$n_{{\hsm}_{({\bbbar})}}$ \\ $\geq 1$}}
 & {\makecell{$n_{{\as}_{(\bbbar)}}$ \\ $\geq 1$}}
 & {\makecell{$m_{_{T}} (p^{({\ell})}_T, \etmiss)$  \\ $> $100 GeV}}
 & {\makecell{$H_T^{(j)}$ \\ $> 300$ GeV}}
 & {\makecell{$R_3$ \\ $< 0.85$}} 
 & {\makecell{$\sigma_{\mathrm{fid}}$ \\ $\times k$}} \\ \hline
    \hline BP1  & 3.64 & 3.13 & 0.91  & 0.17 & 0.04  & 0.029 & 0.027  & 0.0219 & 0.0306   \\ \hline
      \hline BP2  &  4.02 & 3.586 & 1.05  & 0.12 & 0.024  & 0.016 & 0.015  & 0.0103 &  0.0144 \\ \hline
      \hline BP3  &  0.281  & 0.263 & 0.079  & 0.016 & 0.0046  & 0.0036 & 0.0035  &0.0027 & 0.0038  \\ \hline
      \hline BP4 &  0.202 & 0.192 & 0.0585  & 0.007 & 0.0017  & 0.0014 & 0.0012  & 0.0009 & 0.0013  \\ \hline
      \hline BP5 &  21.86 & 18.42 & 5.16  & 0.54 & 0.068  & 0.049 & 0.042  & 0.031 & 0.0434  \\
\hline
\hline
  & \multicolumn{9}{c|}{} \\
 SMBG & \multicolumn{9}{c|}{} \\
\hline
\hline
$\ttbar$ & 510000 & 36104.3 & 16373.1  & 62.4 & 0.29 & 0.166 & 0.104  & 0.0052 & 0.0073 \\
\hline
\hline
$\ttbar \hsm$  & 405 & 46.13 & 22.32 & 1.25 & 0.013  & 0.0044 & 0.0015 & 0.0006 & 0.0007 \\
\hline
\hline
$\ttbar Z$  & 590  & 67.2 & 32.91   & 0.4475 & 0.0068 & 0.0021 & 0.001 & 0.0005 & 0.0006 \\
\hline
\hline
$\ttbar \bbbar$  & 13500  & 879.8 & 391.3  & 5.6 & 0.89 & 0.0068 & -- & --  & -- \\
\hline
\hline 
$\ttbar W^\pm$   & 345  & 51.2 & 25.2 & 0.2 & 0.0006 & 0.0001 & 0.0001 & -- & -- \\
\hline
\end{tabular}
}}
\caption{
Values of the original, cut-flow (both at LO) and the fiducial (after folding in the appropriate
$k$-factors) cross sections (in femtobarns) for various signal scenarios and for the primary SM background (SMBG) processes. 
The markings `$-$' stand for cross-section values that are too small to be mentioned.}
\label{tab:cross-section}
\end{center}
\end{table}
%
%
%
\begin{table}[t]
\centering
 	\begin{tabular}{cccccc}
        \hline 
        Integrated & \multicolumn{5}{c}{Signal significance $\sigma \left(=\frac{S}{\sqrt{S+B}} \right)$} \\ luminosity & BP1 & BP2 & BP3 & BP4 & BP5  \\
        ${\cal L}$ \, (\fbinv) & & & & & \\
        \hline 
300 & 2.7 & 2.0 & 0.6 & 0.2 & 3.3  \\ \hline
3000 & 8.5 & 6.4 & 1.9 & 0.7 & 10.4 \\ \hline
\hline
\end{tabular}
\caption{
Values of projected signal significances under a cut-based analysis for various 
benchmark scenarios and for two values of integrated luminosity at the 13 TeV 
LHC.
}
\label{tab:signalsignificances}
\end{table}

Note that, in the case of BP1 and BP3 where $m_{\widetilde{Q}_3}$ is not so large and the mixings between the `chiral' partners in both the sbottom and the top squark sectors are small, both
$m_{\tilde{b}_1 (\sim \tilde{b}_L)}$ and $m_{\tilde{t}_1 (\sim \tilde{t}_L)}$  would be on the smaller side and close by and hence their pair-production cross sections at the LHC would be of comparable  strengths. Thus, given that the process $pp \to \tilde{b}_1 \tilde{b}_1^*$ could also yield the signal final state of our interest, the corresponding contribution to the same is also to be considered as well. However, this is expected to be suppressed given that $\tilde{b}_1 \sim \tilde{b}_L$ would undergo the dominant decay
$\tilde{b}_1 \rightarrow t \chi_1^\pm$ (with branching ratio $\sim 95$\% ) when $\chi_1^\pm$ is higgsino-like which would not lead to an SM Higgs boson, that we require in our signal, further down the cascade. Nonetheless, there will still be some residual contributions in the signal region we propose that could come from decaying
$W^\pm$-bosons arising in the cascades of the top quarks and the lighter chargino. 
Our simulations for the benchmark points BP1 and BP3 show that the sensitivity of the proposed signal can be increased by up to about 25\%  when compared to the case where, as presented in Table~\ref{tab:signalsignificances}, contributions from only $pp \to \tilde{t}_1 \tilde{t}_1^*$ are considered.

\subsection{A multivariate analysis}
\label{subsec:mva}
The cut-based approach presented in section~\ref{subsec:cut-based} is found to be reasonably efficient in suppressing an otherwise large SM background. However, the signal rate itself being tiny, especially for larger $\mstone$, the LHC experiments hardly become sensitive to the same, even with an integrated luminosity of 3000~\fbinv. Looking for a more optimal set of cuts to mitigate the
problem becomes increasingly difficult when these are to be chosen solely based
on intuition. A smarter set of cuts that segregates the signal events from the background ones in a more efficient manner would thus inevitably refer to complex kinematic entities, which are something that an MVA could essentially sniff out for us in the form of a single `discriminator'.

Towards this, the formalism of BDT (the so-called gradient boosting
technique~\cite{Hocker:2007ht}, to be precise) that we use within the MVA framework creates 
different disjoint decision trees by optimizing the cuts on various kinematic 
variables used for training the BDT and then applying them to subsets of events 
with different signal purities. If a signal event is wrongly classified as a 
background one or vice versa, it `boosts' that event by increasing the 
corresponding weight and initiates a fresh tree with the new weight. Repeating this 
process optimizes the performance of the BDT till it finds the coveted 
discriminator. We use the package Toolkit for Multi Variate Analysis ({\tt TMVA})~\cite{Hocker:2007ht, Voss:2007jxm},
built-in in the CERN {\tt ROOT}~\cite{Brun:1997pa} analysis framework, for this purpose.

We note that though we have some representative values of $\mas$ for our 
benchmark scenarios, this is an unknown parameter from the experimental 
perspective. Hence, in the MVA, choosing a pre-selected mass window to unearth 
a possible light Higgs bosons can be seen as biasing the analysis to a 
specific signal region with a precise kinematic attribute.
On the other hand, leptons and the SM Higgs boson are, by now, rather
well-understood objects at the LHC. Thus, for the MVA, we move away from the
restricted signal region of the cut-based analysis that is characterized by
an $m_{_{\mathrm{FJ} {(bb)}}}$ window about a representative $\mas$ value and require only the following known objects in the final state while training the BDT:
\renewcommand{\theenumi}{\roman{enumi}}%
\begin{enumerate}
\item an isolated lepton, which can effectively suppress the QCD background,
\item an SM Higgs boson is reconstructed in an 
appropriate $\bbbar$ pair within the invariant mass-window
$|m_{_{\mathrm{FJ} {(bb)}}} - \mhsm| < 25$ GeV, as has been already specified in 
section~\ref{subsec:cut-based} and
\item an accompanying $\rm \etmiss > 80$~GeV.
\end{enumerate}
We take the mass 
window of a possible light Higgs boson jet as a free variable 
($m_{j_{_{bb}}} \equiv m_{\text{FJ}(bb)}({\text {low}}) <90$ GeV) without any loss of 
generality.

Once the BDT finds out a suitable discriminator modulo a particular reference to the possible presence of a light Higgs boson like $\as$), we focus on the above mentioned region by putting a cut on the BDT discriminator. As is expected, the resulting sample would primarily contain signal-enriched events, but, depending on the signal purity, would also include some events from the background.
We then look for a light Higgs boson in this filtered sample which is expected to show up as a peak in the 
$m_{j_{_{bb}}}$ distribution.

To train the BDT we use 60$\%$ of the events from each of the signal and the 
background samples, while the other 40\% of the events in each category are 
deployed for the purpose of testing the performance of the BDT.
To start with, a large number of kinematic variables are constructed from 
various low-level objects present in the signal final state. However, on 
learning from the ranking and correlations among these variables as reported 
in the MVA training output, this number is gradually brought down 
to 14. The list of input kinematic variables according to their 
rankings, that tell about their effectiveness in decreasing order, is 
presented in table~\ref{tab:rank_BP1_hsm} for the benchmark scenario BP1.
\begin{table}[t]
\renewcommand{\arraystretch}{1.25}
	\centering
 \begin{tabular}{c c c}
\hline
			Rankings & Variables  & Descriptions\\
			\hline
			1 & $ \Delta R(\ell, \MET)$ & $\Delta R$ between the lepton and $\MET$ in the azimuthal plane \\
			2 & $p^{(\ell)}_{T}$   &	$p_T$ of the lepton \\
			3 & $H_T^{(j)}$      	&	Scalar sum of $p_T$ of all jets \\
			4 &  $N_b$ 	&	Number of $b$-jets \\
            5 & $m_{j_{_{bb}}} \equiv m_{\text{FJ}(bb)}({\text {low}})$  & Jet mass of the $bb$ fat jet with $m_{{\mathrm{FJ {(bb)}}}}$$\rm <90~GeV$\\
			6 & $\Delta R(\ell,\hsm)$ & $\Delta R$ between the lepton and $\hsm$ in the azimuthal plane \\
			7 & $\Delta R(\hsm, \etmiss)$ & $\Delta R$ between $\hsm$ and $\etmiss$ \\
			8 & $p^{(\hsm)}_{T}$   &	$p_T$ of $\hsm$  \\
			9 &  $m_{_{T}} (p^{({\ell})}_T, \etmiss)$ &  Transverse mass of the system containing the lepton and $\etmiss$ \\
			10 & $ \MET$     	&		Missing transverse energy in an event\\
			11 & $R_3$ & The variable `$R_{n=3}$', as defined in section \ref{subsec:cut-based} \\
			12 & $\Delta R(j_1, \etmiss)$ & $\Delta R$ between the leading jet and $\etmiss$ \\
			13 & $N_\mathrm{jet}$ & Number of ordinary jets  \\
			14 & $\Delta R(\ell,j_1)$ &  $\Delta R$ between the lepton and the leading jet \\
			
	\hline
		\end{tabular}
  	\caption{Rankings of various kinematic variables used for training the BDT in the scenario BP1.}
	\label{tab:rank_BP1_hsm}
\end{table}
Since the objects in the signal final state originate in the cascades of 
massive top squarks ($\mstone \sim$ 1 TeV), events possessing a larger number 
of harder jets are predominant for the signal when compared to the ones 
from the background. Hence, as can be gleaned from table~\ref{tab:rank_BP1_hsm}, 
variables sensitive to jet-multiplicity and hardness of jets generally receive 
higher rankings.
Interestingly enough, one can find from table~\ref{tab:rank_BP1_hsm} that various isolation variables ($\Delta R$) generally tend to become important when compared to some prominent variables used in the cut-based analysis.
The relatively higher importance of $\Delta R({\ell, \etmiss})$ and $p_T^{(\ell)}$ have pushed $m_{_{T}} (p^{({\ell})}_T, \etmiss)$, which has 
a good correlation with these two variables, lower down the ranking. Also, once we require an $\hsm$ to be reconstructed in a $b$-jet pair, the dominant background from $\ttbar$ production
will get eliminated
while there may still be a number of $b$-jets left in a signal event. This attaches a reasonable importance to the $b$-jet multiplicity ($N_b$) in an event as a means to filter out the signal. Note that, for all the benchmark scenarios, we have used 
the same set of kinematic variables as inputs to the MVA. 
This is in view of the fact that, although we have presented 
table~\ref{tab:rank_BP1_hsm} for scenario BP1, we find that the 
relative rankings of these kinematic variables remain grossly unaltered across 
the scenarios we are dealing with.

\begin{figure}[t]
\begin{center}
\includegraphics[width=0.48\linewidth]{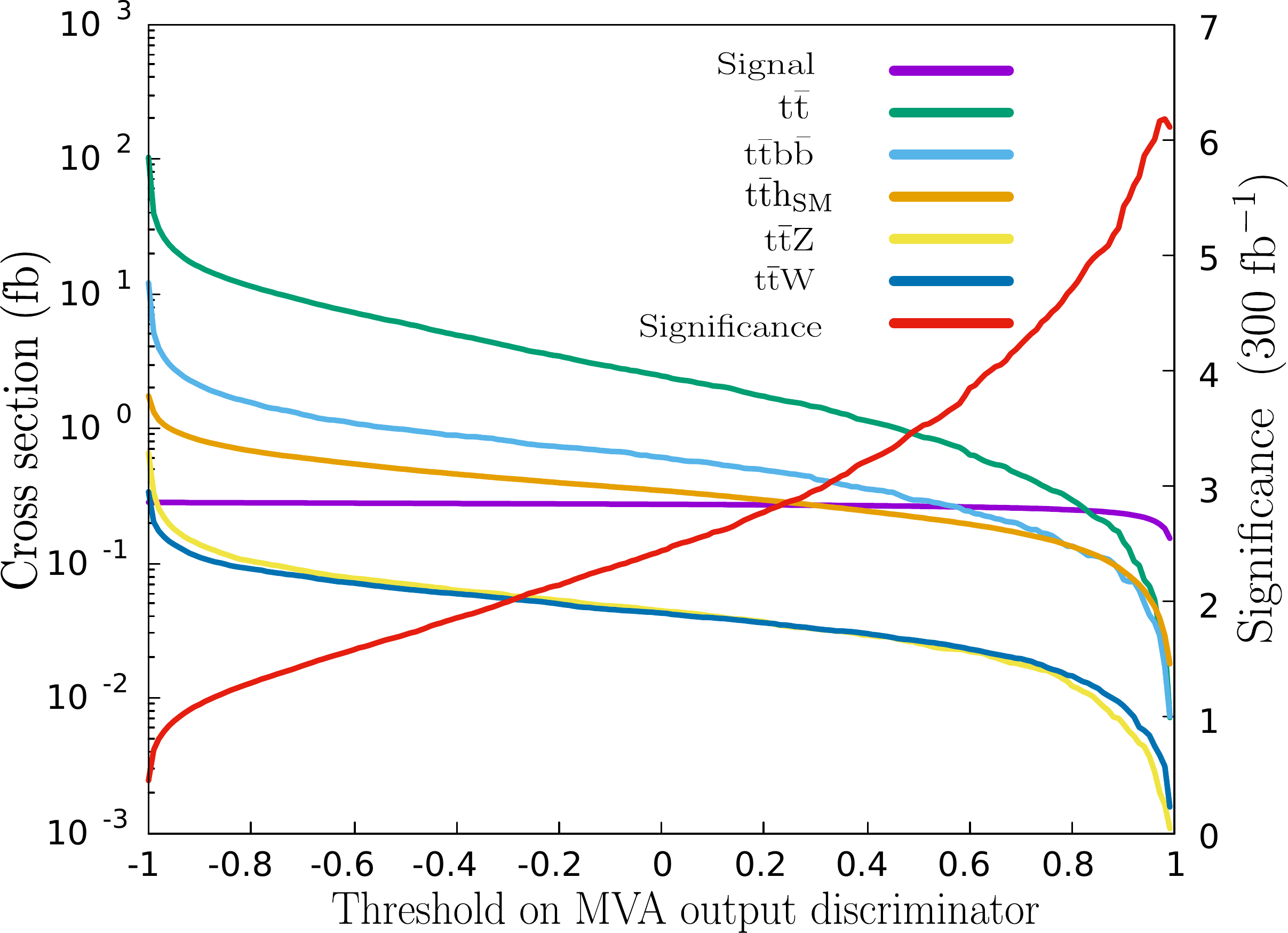}
\hskip 10pt
\includegraphics[width=0.48\linewidth]{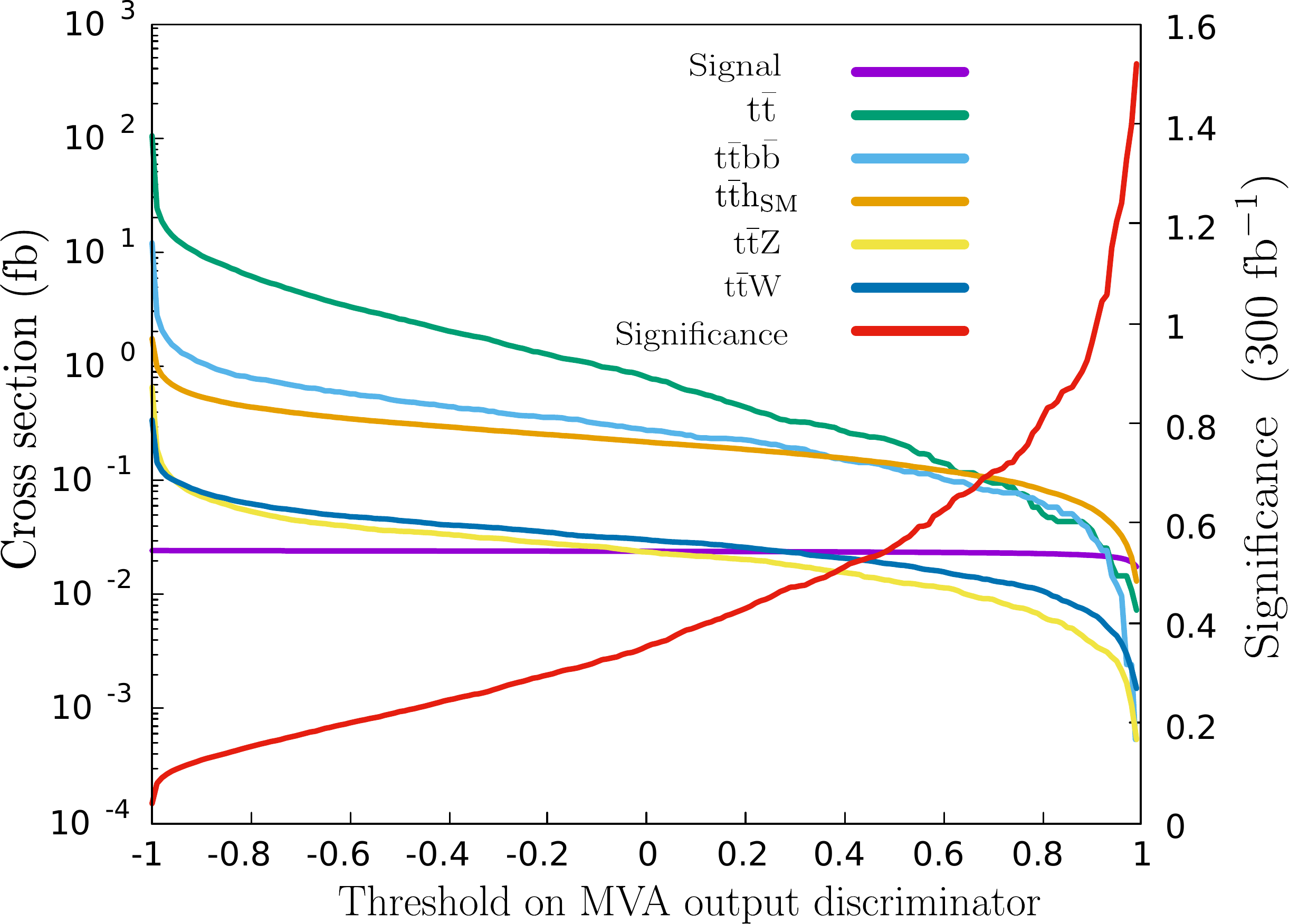}
\caption{ \small Distributions of the MVA output discriminators for the signal and the backgrounds along with the signal significances corresponding to an integrated
luminosity of 300 \fbinv for the benchmark scenario BP1 (left) and BP3 (right).}
\label{fig:mva-sensitivity}
\end{center}
\vspace{-0.5cm}
\end{figure}

Finally, the signal significance for each scenario is evaluated by applying a cut to the 
BDT discriminator. For illustrative purposes, in the left (right) plot of figure~\ref{fig:mva-sensitivity}, 
the yields in scenario BP1 (BP3) and those for the backgrounds are presented 
along with the signal significance as functions of the cut on the MVA output 
discriminator for an integrated luminosity of 300~\fbinv. The plot on 
the left indicates that a signal significance of $\sim 1-6 \, \sigma$ can be 
obtained in scenario BP1 by choosing a lower cut of 0.9 on the MVA output 
discriminator which roughly translates to $\sim 5-20 \, \sigma$ for an integrated luminosity of 3000~\fbinv.
No essential changes in the features are noted
 for the benchmark scenario BP3 except for a general drop in the signal significance, as can be gleaned from  the right plot of figure~\ref{fig:mva-sensitivity}.

%
%
The projected values of signal significance for all the benchmark scenarios are 
presented in table~\ref{tab:sensitivity_mva} corresponding to accumulated luminosities of 300~\fbinv~and 3000 \fbinv.
%
\begin{table}[t]
 	\centering
 	\begin{tabular}{cccccc}
        \hline 
        Integrated & \multicolumn{5}{c}{Signal significance $\sigma \left(=\frac{S}{\sqrt{S+B}} \right)$} \\ luminosity & BP1 & BP2 & BP3 & BP4 & BP5 \\ 
        ${\cal L}$ \, (\fbinv) & & & & & \\ 
        \hline 
    $\rm 300$ & 5.5 &  4.5  & 1.2  & 0.7 & 9 \\ 
    $\rm 3000$& 17 &  14 & 4  & 2.2 & 28 \\ 
\hline
\end{tabular}
\caption{
Values of projected signal significance under MVA for various 
benchmark scenarios and for two values of integrated luminosity at the 13 TeV 
LHC.
}
\label{tab:sensitivity_mva}
\end{table}
%
It is to be noted that a two to three-fold enhancement in the signal 
significance is observed when using the MVA as compared to the cut-based 
approach. This can be attributed to the fact that a multi-dimensional 
treatment of the kinematic cuts by taking the correlations among the kinematic 
variables into account helps MVA arrive at a robust set of selection criteria 
almost without affecting the signal strength. In contrast, in the cut-based 
approach, many of the variables remain practically unused in order to retain 
an optimal number of detectable signal events for a given value of accumulated 
luminosity.
A reasonable signal significance 
of up to $\sim 6\sigma$ is now seen to be achieved in scenario BP1 for an accumulated luminosity of 300~\fbinv~while, for scenario BP3, the significance still remains sub-optimal at the level of~$\sim$1.5$\sigma$.
The signal significance is expected to increase further by a factor of around 3 for an accumulated luminosity of 3000~\fbinv, as the same  scales roughly as $\sqrt{\cal{L}}$.
%

%
\begin{figure}[t]
\centering
\includegraphics[width=8 cm]{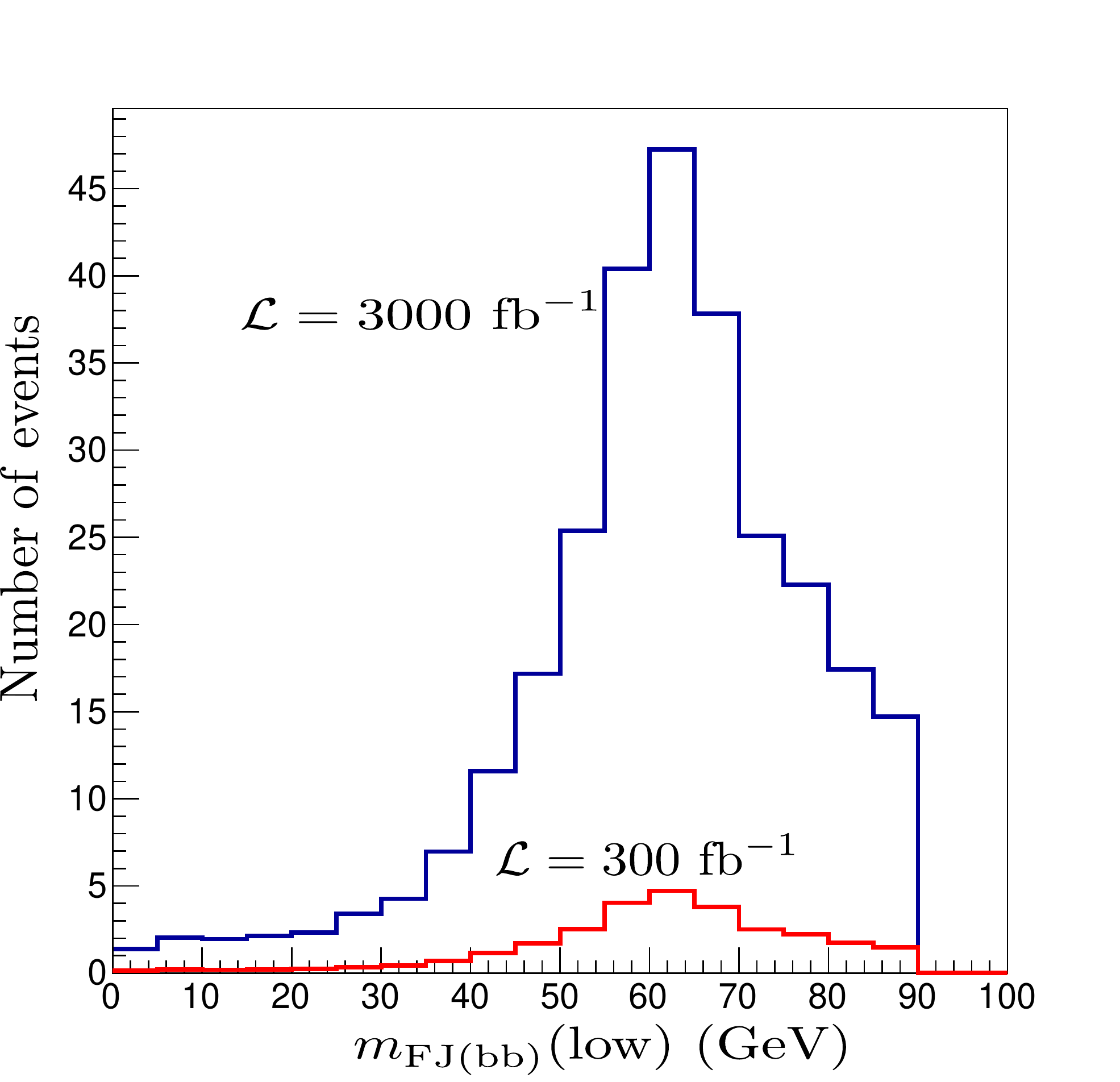}
\caption{
Differential distribution of the number of events in
$m_{_{\mathrm{FJ}_{(bb)}}}$(low) for scenario BP1 with integrated luminosities of 
300~\fbinv~and 3000~\fbinv~and with a cut on the BDT discriminator of 0.9 while
keeping events in which at least one fat jet containing a pair of $b$'s appears.
}
\label{fig:mjbb}
\end{figure}
%
 
A peak in the $m_{j_{_{bb}}}$ distribution would indicate the presence of a light Higgs boson of a possible NMSSM origin could now be searched for in the event sample.
In figure~\ref{fig:mjbb}, the combined 
distribution of $m_{j_{_{bb}}}$ is plotted with the event-sample that survives 
a cut of 0.9 on the BDT  discriminator for the scenario 
BP1.
The clear peak around $m_{j_{_{bb}}}=65$ GeV confirms the presence of a light 
Higgs boson resonance in the signal region uncovered by the MVA. Note that the 
information on the mass of a light Higgs boson has nowhere been directly used 
in this analysis. Thus, the approach should work for any other scenario with a
different assumption on $\mas$ and would be able to find a low mass peak in
the data in case such a resonance is present.

Before we end this section, the following general observations on the projected signal significance, based on a few benchmark scenarios that we consider, can be made:
\begin{itemize}
\item
The cut-based analysis never delivers  a signal significance close to the coveted $5\sigma$ level with 300~\fbinv~worth of LHC data.
\item
Hardly ever a $5\sigma$  significance could be expected with
$\{\mstone, \, \mueff\} \approx \{1.5 \, \mathrm{TeV}, \, 900 \, \mathrm{GeV}\}$, even by employing an MVA and, that also, with data worth 3000~\fbinv.
\item
With 3000~\fbinv~of data, one could expect an $\sim 8\sigma$ ($\sim 6\sigma$) signal significance for ${\stone} \approx \stleft (\stright)$ and for $\{\mstone, \, \mueff\} \approx \{1 \, \mathrm{TeV}, \, 650 \, \mathrm{GeV}\}$ which drops sharply as $\mstone$ and $\mueff$ increase.  
\item
An MVA could deliver a $\gtrsim 5\sigma$ signal significance for $\{\mstone, \, \mueff\} \approx \{1 \, \mathrm{TeV}, \, 650 \, \mathrm{GeV}\}$, irrespective of the chiral content of $\stone$, even with 300~\fbinv~of LHC data, something that can be seen as a rather meaningful improvement over what a cut-based analysis
could offer with the same volume of data.
\end{itemize}
%
\section{Summary and outlook}
\label{sec:summary}
Searches for relatively light ewinos and a top squark have remained to 
be two priority programmes at the LHC, as these are intimately connected to 
 the so-called `naturalness' of a SUSY scenario. Stringent lower bounds that the LHC 
experiments have set on their masses are often based on some simplified theoretical 
assumptions. Even though due regard is to be paid to these emerging constraints 
which might be carrying a general message, departures from those simplistic 
assumptions are known to affect (relax) the reported bounds, sometimes 
drastically, resulting in the possibility that some such new physics 
excitations might have already popped up but are missed by the experiments. 
In this work, we exploit some such possibilities that are characteristics of a
scenario like the popular \z3nmssm and are able to relax the reported lower
bounds on the masses of both the lighter ewinos and $\stone$. We then demonstrate to what extent such ewinos could further 
be accessible in the cascade decays of $\stone$'s, produced in pairs at the LHC, when the masses 
of the ewinos in context exceed the (relaxed) bounds obtained by reinterpreting the results on their direct 
searches at the LHC.

The specific \z3nmssm scenario that we adopt possesses a bino-like LSP and a singlet-like scalar, both 
lighter than 100 GeV, along with a singlino-like NLSP with its mass not much above 100 GeV. 
A pair of immediately heavier neutralinos and the lighter chargino are all higgsino-like with masses around 
or below $\sim 1$ TeV (down to $\sim$500~GeV) while the heaviest neutralino and the heavier chargino are all wino-like and are taken to be rather massive such that those are practically decoupled from the 
physics we are interested in.
The interesting region of the parameter space, in terms of the quantities like $\mone$, $\mueff$, $\msinglino$, $\lambda$, is such that these conspire to give rise to a new 
blind spot (discussed recently in the literature) which leaves the DMDD-SI rate below its current experimental upper bound. On the other hand, the singlet-like scalar offers an annihilation funnel for a
bino-like DM pair thus rendering the relic abundance in the right ballpark, as suggested by relevant experiments.

In addition to being a scenario which is much motivated as it may still offer a bino-like DM (unlike in the
MSSM) satisfying all current constraints, it has the ingredients to trigger SFOEWPT in the early Universe 
thus setting the stage for the coveted EWBG. The relatively light singlet-like scalar serves as a crucial 
excitation for the scenario to work and its presence affects the collider phenomenology in an essential way 
resulting in altered LHC bounds on the masses of the ewinos and the lighter top squark. In the process, the 
scenario offers new signatures to search for at the LHC in the cascade decays of $\stone$'s via which hunt for the lighter of the ewinos could get extended beyond their direct LHC reaches.

The signal final state we study originates in the cascade decays of a pair of $\stone$ that involve the two higgsino-dominated neutralinos, $\ntrlthreefour$, the singlino-dominated NLSP, $\ntrltwo$, and the
singlet-like scalar, $\as$, and is comprised of the parton-level states that are encircled above and, thus, and is given by `$1$ lepton + 4 $b$-jets + $\geq 4 \, \mathrm{jets} + \etmiss$', where the four $b$-jets are required to be reconstructed into an $\hsm$ and an $\as$. The pair-produced $\stone$'s may both decay via the first cascade mode $\ntrlthreefour$ or one of them could  decay via $\charonepm$. As can be seen, such a signal final state is rich in jets, in particular, in $b$-jets, and the sources of the lone lepton and the ordinary jets are the $W^\pm$ bosons appearing in the cascades. Further, both $\hsm$ and $\as$ could be reasonably boosted because of large mass-gaps among $\stone$ and the ewinos appearing in its cascades. Hence, in our analysis, we treat those $b$-jets as fat jets. To summarize, the optimal signal region that emerges from our analysis is characterized by 1 isolated lepton, at least one each of $\hsm$ and $\as$, at least four other hard jets, and a large $\etmiss$.

It is to be noted that the LHC-sensitivity of the proposed final state crucially depends on two decay 
branching fractions, viz., BR[$\ntrlthreefour \to \hsm \ntrltwo$], which is required to dominate over
BR[$\ntrlthreefour \to \hsm \ntrlone$], and BR[$\ntrltwo \to \ntrlone \as$], which should be large.
The first requirement is ensured when `$\lambda$' is on the larger side. Compliance with the second one can 
be achieved for an optimally small mass-split between $\ntrltwo$ and $\ntrlone$ such that the only allowed 
two-body decay of $\ntrltwo$ is $\ntrltwo \to \ntrlone \as$. That  these simultaneous requirements happen to prefer the region of the parameter space that
facilitates an SFOEWPT and that such a scenario having a bino-dominated LSP nicely complies with the constraints from the DM and cosmology experiments,
make the proposed signal all the more characteristic of such a scenario.

We have presented the comparative sensitivities of the 13 TeV LHC to the proposed 
signal (in terms of the signal significance) for accumulated luminosities of
300~\fbinv~and 3000~\fbinv~by referring to a few benchmark scenarios and by 
performing a usual cut-based analysis, followed by a multivariate one. The 
benchmark scenarios are broadly categorized by the masses and the chiral 
contents of $\stone$. While the experimental sensitivity to the signal has the 
expected inverse dependence on $\mstone$, the same on the chiral content of
$\stone$ mounts from the fact that its effective branching fraction to $\hsm$ 
is expected to be significantly larger for $\stone \approx \stleft$. 

We find that, with data worth 300~\fbinv, a usual cut-based analysis would 
not be able to reach out to these ewinos in the cascades of a pair of 
top squarks produced directly at the LHC. Instead, an MVA can be sensitive to 
the higgsino-like ewinos with masses $\gtrsim 650$ GeV when $\stone$ is not much heavier than 1 TeV. On the other hand, with an 
accumulated luminosity of $\sim 3000$ \fbinv, states as massive as the above ones can 
already be probed in a cut-based approach. However, even an MVA on such a data 
set is unlikely to find these ewinos when their masses approach a TeV 
and the $\mstone$ hits $\sim 1.5$~TeV.

It can thus be concluded that with a larger volume of available LHC data and
when aided by the power of the MVA, one should be able to probe significantly heavier higgsino-like ewinos (and hence, larger $\mueff$) of the \z3nmssm scenario at the $\sim 13$ TeV LHC in the cascades of pair-produced $\stone$'s.
However, such a possibility gets quickly restricted as $\mstone$ increases.
Our present analysis tends to indicate that a limiting $~5\sigma$ reach, under 
the most favorable of the setups discussed in this work, could perhaps only be achieved
for $\mstone, \, \mueff \lesssim 1.3 \, \mathrm{TeV}, \, 1 \, \mathrm{TeV}$.
Applications of deep-learning methods might be an obvious way forward to extend the reach in these masses.

Nonetheless, the signal we study remains to be
rather typical to a  spectrum discussed in this work which is both motivated and consolidated from multiple theoretical and experimental  considerations. Hence, once observed, a signal like this could promptly point back to such a spectrum that includes a relatively light ewino and a light scalar, both having a singlet origin, along with a little heavier higgsino-like ewinos and not so heavy
a top squark.
It may, however, be noted that a flipped hierarchy between the bino- and the 
singlino-like states (i.e., the LSP (NLSP) is singlino (bino)-dominated) relative to the scenario we discuss in this work, could also lead to a similar 
final state at the LHC if `$\lambda$' is on the smaller side when
BR[$\ntrlthreefour \to \hsm \ntrltwo$] would again dominate over 
BR[$\ntrlthreefour \to \hsm \ntrlone$].
However, examining the viability of such a scenario in detail in the present context and then looking for a resolution
of a possible ambiguity that it could give rise to are beyond the scope of the present study.
%
%
\section{Acknowledgments}
SR is supported by the funding available from the Department of Atomic Energy (DAE), 
Government of India for the Regional Centre for Accelerator-based Particle Physics 
(RECAPP) at Harish-Chandra Research Institute (HRI). SR is also supported by the Infosys 
award for excellence in research through HRI. SR would like to thank MG and the 
Department of High Energy Physics, Tata Institute of Fundamental Research, Mumbai, India 
for hosting him during the course of this collaborative work.
SR further acknowledges the use of the High-performance Scientific Computing facility 
and RECAPP’s cluster computing facility at HRI  and thanks Chandan Kanaujiya and Ravindra 
Yadav for their technical assistance at these facilities.
%
%

\end{document}